\pdfoutput=1
\documentclass[aps,prd,floats,nofootinbib,preprintnumbers,12pt]{revtex4}
\usepackage{epsfig}
\usepackage{amsmath}
\usepackage{amssymb}
\usepackage{longtable}
\usepackage{array}
\usepackage{color}
\usepackage{graphicx}

\newcommand{\be}{\begin{equation}}
\newcommand{\ee}{\end{equation}}
\newcommand{\bea}{\begin{eqnarray}}
\newcommand{\eea}{\end{eqnarray}}
\newcommand{\ba}{\begin{eqnarray}}
\newcommand{\ea}{\end{eqnarray}}
\newcommand{\s}{\scriptscriptstyle}

\begin{document}

\preprint{FTUV/09-0720}
\preprint{IFIC/09-33}
\preprint{LA-UR 09-03949}

\vspace{3cm}

\title{\bf
Semileptonic decays of light quarks
beyond the Standard Model}
%
%
\author{Vincenzo Cirigliano}\email{cirigliano@lanl.gov}
\affiliation{Theoretical Division,
Los Alamos National Laboratory, Los Alamos, NM 87545}
\author{Mart\'{\i}n Gonz\'alez-Alonso}\email{martin.gonzalez@ific.uv.es}
\affiliation{Theoretical Division,
Los Alamos National Laboratory, Los Alamos, NM 87545}
\affiliation{Departament de F\'{\i}sica Te\`orica and IFIC, Universitat de Val\`encia-CSIC,\\
Apt. Correus 22085, E-46071 Val\`encia, Spain}
\author{James P.\ Jenkins}\email
{jjenkins6@lanl.gov}
\affiliation{Theoretical Division,
Los Alamos National Laboratory, Los Alamos, NM 87545}

\begin{abstract}
We  describe  non-standard contributions to semileptonic processes in a model independent way
in terms of  an  $SU(2)_L \times U(1)_Y$ invariant effective lagrangian at the weak scale,
from which we  derive the low-energy effective lagrangian governing muon  and beta decays.
We find that the deviation from  Cabibbo universality,  $\Delta_{\rm CKM} \equiv
|V_{ud}|^2 + |V_{us}|^2 + |V _{ub}|^2 - 1$,  receives contributions from
four effective operators.
The phenomenological bound  $\Delta_{\rm CKM} = (-1 \pm 6) \times 10^{-4}$ provides 
strong constraints on  all four operators, 
corresponding to an effective scale $\Lambda > 11$ TeV (90\% CL).
Depending on the operator,  
this constraint is 
at the same level or better then the Z pole observables. 
Conversely,  precision electroweak constraints  alone would  allow
universality violations as large as 
$\Delta_{\rm CKM}=-0.01$  (90\% CL).
An observed  $\Delta_{\rm CKM} \neq 0$  at this level could  be explained in terms of
a single  four-fermion operator
which is relatively poorly constrained by electroweak  precision measurements.

\end{abstract}

\maketitle

\section{Introduction}\label{sec:Introduction}

Precise lifetime and branching ratio measurements~\cite{Amsler:2008zzb}
combined with improved theoretical control of hadronic matrix
elements and radiative corrections  make
semileptonic  decays of light quarks
a deep probe of the nature of weak interactions~\cite{Marciano:2007zz,Antonelli:2008jg}.
In particular, the determination of the  elements $V_{ud}$ and $V_{us}$
of the  Cabibbo-Kobayashi-Maskawa (CKM)~\cite{Cabibbo:1963yz,Kobayashi:1973fv}
quark mixing matrix is  approaching the  0.025\% and 0.5\% level, respectively.
Such precise knowledge of $V_{ud}$ and $V_{us}$  enables  tests of Cabibbo  universality,
equivalent to  the  CKM  unitarity condition\footnote{$V_{ub} \sim 10^{-3}$
contributes negligibly to this relation.}
$|V_{ud}|^2 + |V_{us}|^2 + |V_{ub}|^2 =1$,
at the level of $0.001$ or better.
Assuming that new physics contributions scale
as $\alpha/\pi   (M_W^2/\Lambda^2)$,
the unitarity test probes energy scales  $\Lambda$ on the order of the TeV,
which will be directly probed at the LHC.

While the consequences of Cabibbo universality tests on Standard Model  (SM) extensions
have been considered in a number of explicit (mostly supersymmetric)
scenarios~\cite{Barbieri:1985ff,Marciano:1987ja,Hagiwara:1995fx,Kurylov:2001zx},
a model-independent analysis  of semileptonic processes beyond the SM is missing.
The goal of this investigation is to  analyze in a model-independent  effective theory setup
new physics  contributions to low energy charged-current (CC) processes.
The resulting framework allows us to assess  in a fairly general way the impact
of semileptonic processes in
constraining and discriminating  SM extensions.
We shall pay  special attention to purely leptonic
and semileptonic decays of light hadrons used to extract the CKM elements
$V_{ud}$ and $V_{us}$.

Assuming the existence of a mass gap between the SM and its extension,
we parameterize  the effect of  new degrees of freedom and interactions beyond the SM
via  a series of higher dimensional operators constructed with the low-energy SM fields.
If the SM extension is weakly coupled,
the resulting TeV-scale effective lagrangian linearly realizes
the electro-weak (EW) symmetry $SU(2)_L \times U(1)_Y$   and  contains
a SM-like  Higgs doublet~\cite{Buchmuller:1985jz}.
This method  is quite general and allows us to study the implications
of precision measurements  on a large class of  models.
In particular,  the  effective theory approach allows us to
understand in a model-independent way
(i)  the significance of Cabibbo universality constraints compared to
other precision measurements
(for example, could we expect sizable deviations from universality
in light of no deviation from the SM in precision tests at the Z pole?);
(ii) the correlations between possible deviations  from universality
and  other precision observables,
not always simple to identify  in  a specific model analysis.

This article is organized as follows.
In Section~\ref{sect:weakscale}
we review the form of the  most general weak scale effective lagrangian including operators
up to dimension six,  contributing to precision electroweak measurements {\it and} semileptonic decays.
In Section~\ref{sect:gevscale}
we derive the low-energy ($O(1)$ GeV)  effective lagrangian describing purely leptonic  and semileptonic
CC interaction.
We discuss the flavor structure of the relevant effective couplings in Section~\ref{sect:flavor}.
In Section~\ref{sect:pheno1}
we give an overview of the phenomenology of $V_{ud}$ and $V_{us}$ beyond the SM,
and derive the relation between universality violations and other precision measurements
at the operator level.
Section~\ref{sec:InvaraintAnalysis}
is devoted to a quantitative analysis of the interplay between the universality constraint
and other precision measurements, while
Section~\ref{sec:Conclusions}
contains our conclusions.

\section{Weak scale effective lagrangian}
\label{sect:weakscale}

As discussed in the introduction, our aim is to
analyze in a model-independent framework   new physics contributions
to both precision electroweak observables and  beta decays.
Given the successes of the SM
at energies up to the  electroweak  scale  $v \sim 100$ GeV,
we adopt here  the point of view that
the SM is the low-energy limit of a more fundamental theory.
Specifically,  we assume that: (i) there is a  gap between the weak scale $v$
and the scale $\Lambda$  where new degrees of freedom appear;
(ii) the SM extension at the weak scale is  weakly coupled,
so the EW symmetry $SU(2)_L \times U(1)_Y$ is linearly realized
and the low-energy theory contains a SM-like  Higgs doublet~\cite{Buchmuller:1985jz}.
Analyses of EW precision data in nonlinear realizations of EW symmetry
can be found in the
literature~\cite{Appelquist:1993ka, Longhitano:1980iz, Feruglio:1992wf,Wudka:1994ny}.
In the spirit of the effective field theory approach, we integrate out all the heavy fields and
describe physics at the weak scale (and below) by means of an
effective non-renormalizable lagrangian  of the form:
\bea
\label{eq:EFT}
{\cal L}^{(\rm{eff})} &=& {\cal L}_{\rm{SM}} + \frac{1}{\Lambda} {\cal L}_5 + \frac{1}{\Lambda^2} {\cal L}_6
 + \frac{1}{\Lambda^3} {\cal L}_7 + \ldots \\
{\cal L}_n &=& \sum_i \alpha^{(n)}_i~ O_i^{(n)}~,
\eea
where $\Lambda$ is the characteristic scale of the new physics
and  ${\cal O}_i^{(n)}$ are local gauge-invariant operators of dimension $n$ built out of SM fields.
Assuming that right-handed neutrinos do not appear as low-energy degrees of
freedom,  the building blocks to construct local operators are
the  gauge fields $G_\mu^A,  \, W_\mu^a, \, B_\mu$,
corresponding to $SU(3)\times SU(2)_L \times U(1)_Y$,
the five  fermionic  gauge multiplets,
\be
l^i =
\left(
\begin{array}{c}
\nu_L^i\\
e_L^i
\end{array}
\right)
\qquad e^i = e_R^i
\qquad q^i =
\left(
\begin{array}{c}
u_L^i \\
d_L^i
\end{array}
\right)  \qquad
u^i = u_R^i \qquad
d^i = d_R^i~,
\label{eq:fermions}
\ee
the Higgs doublet $\varphi$
\be
\varphi =
\left(
\begin{array}{c}
\varphi^+ \\
\varphi^0
\end{array}
\right)~,
\ee
and the covariant derivative
\be
D_{\mu} =   I \, \partial_\mu   \, - \, i g_s \frac{\lambda^A}{2} G_\mu^A \,
- \, i g \frac{\sigma^a}{2} W_\mu^a   \, - \, i g'  Y B_\mu~.
\ee
In the above expression  $\lambda^A$ are the $SU(3)$ Gell-Mann matrices,
$\sigma^a$ are the $SU(2)$  Pauli matrices,  $g_s, g, g'$ are the gauge couplings
and $Y$ is the hypercharge of a given multiplet.

In our analysis  we will  not consider operators that violate total lepton and  baryon number
(we assume they are suppressed by a scale much higher than $\Lambda \sim {\rm TeV}$~\cite{deGouvea:2007xp}).
Under the above assumptions,
it can be shown~\cite{Buchmuller:1985jz} that the first corrections to the SM lagrangian are of dimension six.  A complete set of dimension-six operators is given
in the pioneering work of Buchm\"uller and Wyler (BW)~\cite{Buchmuller:1985jz}\footnote{In the
original list of BW there are eighty operators, but it can be shown that it can be
reduced to seventy-seven (see Appendix \ref{sec:BWcomments}).}.
Truncating the expansion at this order we have
\bea
{\cal L}_{BW}^{(\rm{eff})}
= {\cal L}_{\rm{SM}} + \sum_{i=1}^{77} \frac{\alpha_i}{\Lambda^2}~ O_i~.
\eea
For  operators involving quarks and leptons,
both the coefficients  $\alpha_i$ and the operators $O_i$  carry flavor indices.
When needed, we will make the flavor indices explicit, using the notation $[\alpha_i]_{abcd}$
for four-fermion operators.

The above effective lagrangian allows one to parameterize non-standard corrections
to any observable  involving  SM particles.
The contribution from the dimension six operators involve terms proportional to
$v^2/\Lambda^2$ and $E^2/\Lambda^2$, where $v= \langle \varphi^0 \rangle \simeq174 \,  \rm{GeV}$
is the vacuum expectation value (VEV) of the Higgs field and $E$ is the characteristic energy
scale of a given process. In order to be consistent with the truncation of \eqref{eq:EFT} we will work at linear order in the above ratios.

We are interested in the minimal subset of the BW  basis that contribute  at tree level
to CP-conserving  electroweak precision observables {\it and} beta decays.
Upon imposing these requirements (see Appendix~\ref{sec:BWcomments})
we end up with a  basis involving twenty-five operators.
In selecting the operators, flavor  symmetries played no role (in fact at this level
the coefficients $\alpha_i$ can carry any flavor structure).
However, in order to organize the subsequent  phenomenological analysis, it is
useful to classify the operators  according to their
behavior under the  $U(3)^5$  flavor symmetry of the SM gauge lagrangian
(the freedom to perform $U(3)$ transformations in family space
for each of the five fermionic gauge multiplets, listed in Eq.~\ref{eq:fermions}).

\subsection{$U(3)^5$ invariant operators}
\label{sect:U35invariant}

The operators that contain only vectors and scalars are
\begin{equation}
\label{eq:owbh}
 O_{W\!B}=(\varphi^\dagger \sigma^a \varphi) W^a_{\mu \nu} B^{\mu \nu}, \  \  \   O_\varphi^{(3)} = | \varphi^\dagger D_\mu \varphi|^2~.
\end{equation}
There are eleven four-fermion operators:
\begin{eqnarray} &&
  O_{ll}^{(1)}=\frac{1}{2} (\overline{l} \gamma^\mu l) (\overline{l} \gamma_\mu l),
~~~~
O_{ll}^{(3)} = \frac{1}{2} (\overline{l} \gamma^\mu \sigma^a l) (\overline{l} \gamma_\mu \sigma^a l)      \label{eq:oll} \\ &&
  O_{l q}^{(1)}= (\overline{l} \gamma^\mu l) (\overline{q} \gamma_\mu q), ~~~
  O_{l q}^{(3)}= (\overline{l} \gamma^\mu \sigma^a l) (\overline{q} \gamma_\mu \sigma^a q),
     \label{eq:olq} \\ &&
  O_{le}= (\overline{l} \gamma^\mu l) (\overline{e} \gamma_\mu e),  ~~~
  O_{qe}=(\overline{q} \gamma^\mu q) (\overline{e} \gamma_\mu e),
     \label{eq:olqe}  \\ &&
  O_{lu}= (\overline{l} \gamma^\mu l) (\overline{u} \gamma_\mu u),  ~~~
  O_{ld}= (\overline{l} \gamma^\mu l) (\overline{d} \gamma_\mu d),
      \label{eq:olud} \\ &&
  O_{ee}\!=\!\frac{1}{2} (\overline{e} \gamma^\mu e) (\overline{e} \gamma_\mu e), ~~
  O_{eu}\!=\!(\overline{e} \gamma^\mu e) (\overline{u} \gamma_\mu u),  ~~
  O_{ed}\!=\!(\overline{e} \gamma^\mu e) (\overline{d} \gamma_\mu d).
      \label{eq:oeeud}
  \end{eqnarray}
Some comments are in order.
In principle, in order to avoid redundancy (see discussion in Appendix~\ref{sec:BWcomments}), 
one must discard either $O_{ll}^{(3)}$ or  $O_{ll}^{(1)}$. 
However,  here we have followed the 
common practice 
to work  with both 
operators and consider  
only flavor structures factorized according to fermion bilinears. 
Moreover, we use the structure $\bar{L}\gamma_\mu L\cdot\bar{R}\gamma^\mu R$ in operators
\eqref{eq:olqe}, instead of their Fierz transformed $\bar{L}R\cdot\bar{R}L$, that BW use. They are related by a factor ($-2$).

There are seven operators containing two fermions that alter the couplings of
fermions to the gauge bosons:
\begin{eqnarray} &&
   O_{\varphi l}^{(1)} =\! i (\varphi^\dagger D^\mu \varphi)(\overline{l} \gamma_\mu l) +\!{\rm h.c.}, ~~
   O_{\varphi l}^{(3)}=\! i (h^\dagger D^\mu \sigma^a \varphi)(\overline{l} \gamma_\mu \sigma^a l)+\!{\rm h.c.},
     \label{eq:ohl} \\ &&
   O_{\varphi q}^{(1)} =\! i (\varphi^\dagger D^\mu \varphi)(\overline{q} \gamma_\mu q)+\!{\rm h.c.}, ~
   O_{\varphi q}^{(3)} =\! i (\varphi^\dagger D^\mu \sigma^a \varphi)(\overline{q} \gamma_\mu \sigma^a q)+\!{\rm h.c.},
    \label{eq:ohq} \\ &&
      O_{\varphi u} =\! i (\varphi^\dagger D^\mu \varphi)(\overline{u} \gamma_\mu u)+\!{\rm h.c.}, ~
   O_{\varphi d} =\! i (\varphi^\dagger D^\mu \varphi)(\overline{d} \gamma_\mu d)+\!{\rm h.c.},
           \label{eq:Ohud}  \\ &&
   O_{\varphi e} =\! i (\varphi^\dagger D^\mu \varphi)(\overline{e} \gamma_\mu e)+\!{\rm h.c.}
     \label{eq:ohe}
\end{eqnarray}

Finally, there is one operator that  modifies the triple gauge  boson interactions
 \begin{equation}
 \label{eq:ow}
  O_W = \epsilon^{abc} \, W^{a \nu}_{\mu} W^{b\lambda}_{\nu} W^{c \mu}_{\lambda}.
\end{equation}

\subsection{Non $U(3)^5$ invariant operators}
Three are three four-fermion operators
\begin{eqnarray}
&& O_{qde} = (\overline{\ell} e) (\overline{d} q)+ {\rm h.c.},
\label{eq:oqde} \\
&& O_{l q} = (\bar{l}_a e)\epsilon^{ab}(\bar{q}_b u)+ {\rm h.c.}
\label{eq:olq2}
\ \ \ O^t_{l q} = (\bar{l}_a\sigma^{\mu\nu}e)\epsilon^{ab}(\bar{q}_b\sigma_{\mu\nu}u)+ {\rm h.c.}
\end{eqnarray}
and one operator with two fermions
\begin{eqnarray}
&&  O_{\varphi \varphi} = i(\varphi^T \epsilon  D_\mu \varphi) (\overline{u}\gamma^\mu d)+ {\rm h.c.}~,
\label{eq:ohh}
\end{eqnarray}
which gives rise to a right handed charged current coupling.

The twenty-one  $U(3)^5$ invariant operators contribute to  precision EW measurements
(see Ref.~\cite{Han:2004az}),  whereas only nine
of the twenty-five operators contribute to  the semileptonic decays, including all four
$U(3)^5$ breaking  operators.

We conclude this section with some remarks  on our convention for the
coefficients of the ``flavored'' operators:
(i) in those operators that include the h.c. in their definition, the flavor
matrix $\alpha$ will appear in the h.c.-part with a dagger;
(ii) for the operators $O_{ll}^{(1,3)}$ and $O_{ee}$, because of the symmetry between
the two bilinears,  we impose
$\left[\alpha\right]_{ijkl} = \left[\alpha\right]_{klij}$;
(iii) in order to ensure the hermiticity of the operators \eqref{eq:oll}-\eqref{eq:oeeud}
we impose
$\left[\alpha\right]_{ijkl} = \left[\alpha\right]_{jilk}^*$.
None of these conditions entails any loss of generality.

\section{Effective lagrangian for $\mu$ and  quark  $\beta$  decays}
\label{sect:gevscale}

Our task is to identify new physics contributions to low-energy CC processes.
In order to achieve this goal, we need to derive from the
the effective lagrangian at the weak scale (in which heavy gauge bosons
and heavy fermions are still active degrees of freedom)
a  low-energy effective lagrangian describing muon and quark CC
decays.      The analysis involves several steps which we discuss in some detail,
since a complete derivation is missing in the literature, as far as we know.

\subsection{Choice of weak basis for fermions}
At the level of weak scale effective lagrangian,
we can use the $U(3)^5$ invariance to pick a particular basis for the fermionic fields.
In general, a $U(3)^5$ transformation leaves the gauge part of the lagrangian invariant
while affecting both the Yukawa couplings and the coefficients  $\alpha_i$
of dimension  six operators involving fermions.
We perform a specific  $U(3)^5$ transformation that diagonalizes
the down-quark and charged lepton Yukawa matrices $Y_D$ and $Y_E$
and  puts  the up-type
Yukawa matrix in the form $Y_U = V^\dagger  \, Y_U^{\rm diag}$,
where $V$ is the CKM matrix.
The flavored coefficients $\alpha_i$ correspond to this specific choice of
weak basis for the fermion fields.

\subsection{Electroweak symmetry breaking: transformation to propagating eigenstates}
Once the Higgs acquires a VEV the quadratic part of the lagrangian for gauge bosons and fermions
becomes non-diagonal, receiving contributions from both SM interactions and dimension six operators.
In particular, the NP contributions induce kinetic mixing of the weak gauge bosons, in addition
to the usual mass mixing.
Therefore the next step is to perform a change of basis so  that the new fields have
canonically normalized kinetic term and definite masses.

Let us first discuss the gauge boson sector.
We agree with the BW results on the definition of gauge field mass eigenstates and
on the expressions for  the physical masses (Ref. \cite{Buchmuller:1985jz}, section 4.1).
However,  we find small differences from their results in
the couplings of the $W$ and $Z$ to fermion pairs, which can be
written as (ref. \cite{Buchmuller:1985jz}, section 4.2):
\bea
{\cal L}_J &=& \frac{g}{\sqrt{2}} \left( J^C_\mu W^{+\mu} + h.c. \right) + \frac{g}{\cos{\theta_W^0}} J_\mu^N Z^\mu \\
J_\mu^C
        &=& \bar{\nu}_L \gamma_\mu \eta(\nu_L) e_L + \bar{u}_L \gamma_\mu \eta(u_L) d_L + \bar{u}_R \gamma_\mu \eta(u_R) d_R \\
J_\mu^N &=& \bar{\nu}_L \gamma_\mu \epsilon(\nu_L) \nu_L + \bar{e}_L \gamma_\mu \epsilon(e_L) e_L + \bar{u}_L \gamma_\mu \epsilon(u_L) u_L+ \bar{d}_L \gamma_\mu \epsilon(d_L) d_L \nonumber\\
        & & + \bar{e}_R \gamma_\mu \epsilon(e_R) e_R + \bar{u}_R \gamma_\mu \epsilon(u_R) u_R+ \bar{d}_R \gamma_\mu \epsilon(d_R)d_R~.
\eea
Here the $\epsilon$'s and $\eta$'s are $3\times3$ matrices in flavor space. In the case of the charged current we find (BW do not have the $\dagger$ in $\alpha_{\varphi l}^{(3)}$ and
$\alpha_{\varphi q}^{(3)}$)
\bea
\eta(\nu_L) &=&
 \mathbb{I} + 2 \, \hat{\alpha}_{\varphi l}^{(3)\dagger}  \\
\eta(u_L) &=&
  \mathbb{I} + 2 \, \hat{\alpha}_{\varphi q}^{(3)\dagger}  \\
\eta(u_R) &=& - \hat{\alpha}_{\varphi \varphi}~,
\eea
where we have introduced the notation
\be
\hat{\alpha}_X =  \frac{v^2}{\Lambda^2} \, \alpha_X~.
\ee
In the case of the neutral current ($\epsilon$ coefficients) we obtain the same results
as BW except  for the following replacement:
\bea
\hat{\alpha}_X \to \hat{\alpha}_X + \hat{\alpha}_X^\dagger
\label{eq:BWmistake}
\eea
for $\alpha_X=\alpha_{\varphi l}^{(3)},\alpha_{\varphi l}^{(1)}, \alpha_{\varphi q}^{(3)}, \alpha_{\varphi q}^{(1)}, \alpha_{\varphi e}, \alpha_{\varphi u}, \alpha_{\varphi d}$.

Finally, we need to diagonalize the fermion mass matrices. With our choice of weak
basis for the fermions,   the only step that is left is the diagonalization of
the up-quark  mass matrix,  proportional to the Yukawa matrix $Y_U = V^\dagger Y_U^{\rm diag}$,
where $V$ is the CKM matrix.
This can be accomplished by  a $U(3)$ transformation of
the $u_{L}$ fields:
\be
u_{L} \to
V^\dagger u_{L}~.
\ee
As a consequence, the charged current and neutral current couplings involving
up quarks change as follows:
\bea
\eta(u_L) &\to& V ~\eta(u_L) \nonumber\\
\epsilon(u_L) &\to& V~\epsilon(u_L)~V^\dagger~.
\eea
Similarly,  appropriate insertions of  the CKM matrix will appear
in every operator that contains the $u_{L}$ field.

\subsection{Effective lagrangian for muon decay}

The muon decay amplitude  receives contributions from
gauge boson exchange diagrams  (with modified couplings) and from contact operators
such as  $O_{ll}^{(1)}$, $O_{ll}^{(3)}$, $O_{le}$.
Since we work to first order in $v^2/\Lambda^2$,  we do not  need to consider
diagrams contributing to $\mu \to e \bar{\nu}_\alpha \nu_\beta $ with the ``wrong neutrino flavor'',
because they would  correct the muon decay rate to $O(v^4/\Lambda^4)$.
After integrating out the $W$ and $Z$,   the muon decay effective lagrangian reads:
\be
{\cal L}_{\mu \to e \bar{\nu}_e \nu_\mu}
= \frac{-g^2}{2 m_W^2} \Bigg[ \left(1 +
\tilde{v}_L
\right) \cdot \bar{e}_{\s{L}} \gamma_\mu \nu_{e\s{L}}
\ \bar{\nu}_{\mu \s{L}} \gamma^\mu \mu_{\s{L}}     \ + \
\tilde{s}_R
\cdot \bar{e}_{\s{R}} \nu_{e\s{L}}  \ \bar{\nu}_{\mu \s{L}} \mu_{\s{R}} \Bigg] ~+~ h.c.~, ~
\label{eq:leffmu}
\ee
where $m_W^2 = 1/2 g^2 v^2$ is the uncorrected W mass and
\bea
\label{eq:A}
\tilde{v}_L
&= &
 2~[\hat{\alpha}_{\varphi l}^{(3)}]_{11+22^*} - [\hat{\alpha}_{ll}^{(1)}]_{1221} - 2 [\hat{\alpha}_{ll}^{(3)}]_{1122-\frac{1}{2}(1221)}    \\
\label{eq:B}
\tilde{s}_R
&=& +2 [\hat{\alpha}_{le}]_{2112}~,
\eea
represent the correction to the standard $(V-A)\otimes(V-A)$ structure and the coupling associated
with the new $(S-P)\otimes(S+P)$ structure, respectively.

\subsection{Effective lagrangian for beta decays: $d_j \to u_i \,  \ell^- \,  \bar{\nu}_\ell$}

The low-energy effective lagrangian for semileptonic transitions receives
contributions from both W exchange diagrams (with modified W-fermion couplings)
and the four-fermion operators  $O_{l q}^{(3)}$, $O_{qde}$, $O_{l q}$,  $O_{l q}^t$.
As in the muon case, we neglect lepton flavor violating  contributions (wrong neutrino flavor).
The resulting low-energy effective lagrangian governing semileptonic transitions
 $d_j \to u_i \,  \ell^- \,  \bar{\nu}_\ell$  (for a given lepton flavor $\ell$)  reads:
\bea
{\cal L}_{d_j \to u_i \ell^- \bar{\nu}_\ell}
&=&  \frac{-g^2}{2 m_W^2} \, V_{ij} \, \Bigg[
 \Big(1 + [v_L]_{\ell \ell ij} \Big) \   \bar{\ell}_L \gamma_\mu  \nu_{\ell L}    \ \bar{u}_L^i \gamma^\mu d_L^j
 \ + \  [v_R]_{\ell \ell ij}  \   \bar{\ell}_L \gamma_\mu  \nu_{\ell L}    \ \bar{u}_R^i \gamma^\mu d_R^j
\nonumber\\
&+&  [s_L]_{\ell \ell ij}  \   \bar{\ell}_R   \nu_{\ell L}    \ \bar{u}_R^i  d_L^j
\ + \  [s_R]_{\ell \ell ij}  \   \bar{\ell}_R   \nu_{\ell L}    \ \bar{u}_L^i  d_R^j
\nonumber \\
& + &   [t_L]_{\ell \ell ij}  \   \bar{\ell}_R   \sigma_{\mu \nu} \nu_{\ell L}    \ \bar{u}_R^i   \sigma^{\mu \nu} d_L^j
\Bigg]~+~h.c.~,
\label{eq:leffq}
\eea
where
\bea
\label{eq:beta01}
V_{ij} \cdot
 \left[v_{L}\right]_{\ell \ell i j}
&=& 2 \, V_{ij}  \,  \left[\hat{\alpha}_{\varphi l}^{(3)}\right]_{\ell\ell}   +   2 \, V_{im} \left[\hat{\alpha}_{\varphi q}^{(3)}\right]_{jm}^*
-    2\, V_{im} \left[\hat{\alpha}_{l q}^{(3)}\right]_{\ell\ell mj}   \\
V_{ij} \cdot  \left[v_R\right]_{\ell \ell ij } &=& - \left[\hat{\alpha}_{\varphi \varphi}\right]_{ij} \\
V_{ij} \cdot  \left[s_L\right]_{\ell \ell ij } &=& - \left[\hat{\alpha}_{l q}\right]_{\ell\ell ji}^* \\
V_{ij} \cdot \left[s_R\right]_{\ell \ell ij} &=& -  V_{im}\left[\hat{\alpha}_{qde}\right]_{\ell\ell jm}^*  \\
V_{ij} \cdot  \left[t_L\right]_{\ell \ell ij } &=& - \left[\hat{\alpha}^t_{l q} \right]_{\ell\ell ji}^* ~.
\label{eq:beta1}
\eea
In Eqs.~(\ref{eq:beta01}-\ref{eq:beta1}) the repeated indices $i,j,\ell$ are not summed over, while the index $m$ is.

\section{Flavor structure of the effective couplings}
\label{sect:flavor}

So far we have presented our results for the effective lagrangian
keeping generic flavor structures in the  couplings $[\hat{\alpha}_X]_{abcd}$
(see  Eqs.~\ref{eq:A}, \ref{eq:B}, and \ref{eq:beta01} through \ref{eq:beta1}).
However, some of the operators considered in the analysis contribute to flavor changing neutral
current (FCNC)  processes, so that  their flavor structure cannot be generic
if the effective scale is around $\Lambda \sim {\rm TeV}$: the off-diagonal coefficients are
experimentally constrained to be very small.
While it is certainly possible that some operators (weakly  constrained by FCNC) have
generic structures,
we would  like to understand the  FCNC suppression needed for many operators  in terms  of a symmetry principle.
Therefore, we organize the discussion in terms of  perturbations around the $U(3)^5$  flavor symmetry limit.

If the underlying new physics respects the
$U(3)^5$  flavor symmetry of the SM gauge lagrangian,
no problem arises from FCNC constraints.
The largest contributions to the  coefficients are flavor conserving and universal.
Flavor breaking contributions
arise through SM radiative corrections,
due to insertions of Yukawa matrices that break the $U(3)^5$ symmetry.
As a consequence, imposing exact $U(3)^5$ symmetry on the underlying model does not seem realistic.
A weaker assumption, the Minimal Flavor Violation (MFV) hypothesis,  requires
that  $U(3)^5$ is broken in the underlying model only by structures proportional to the SM Yukawa
couplings~\cite{Chivukula:1987py,Hall:1990ac,Buras:2000dm,D'Ambrosio:2002ex},
and by the structures generating neutrino masses~\cite{Cirigliano:2005ck}.
We will therefore organize our discussion in several stages:
\begin{itemize}
\item[1.]  first, assume dominance of $U(3)^5$ invariant operators;
\item[2.] consider effect of $U(3)^5$ breaking induced within MFV;
\item[3.]  consider the effect of generic non-MFV flavor structures.
\end{itemize}

In order to proceed with this program,
for the relevant operators  we list below the flavor structures allowed within MFV.
The notation is as follows:
we denote by  $\bar{\lambda}_{u,d,e}$ the diagonal Yukawa matrices;
$\bar{m}_\nu$ represents the diagonal light neutrino mass matrix;
$V$ denotes the CKM matrix, while $U$ is the PMNS~\cite{Maki:1962mu}
neutrino mixing matrix;
$v$ is the Higgs VEV and $\Lambda_{LN}$ is the scale of lepton number violation,
that appears in the definition of MFV in the lepton sector (we follow here the ``minimal''  scenario
of  Ref.~\cite{Cirigliano:2005ck}).   With this notation,
the leading ``left-left''   flavor structures in the quark and lepton sector read:
\bea
\Delta_{LL}^{(q)} &=&  V^\dagger \,  \bar{\lambda}_u^2 \, V \\
\Delta_{LL}^{(\ell)} &=&  \frac{\Lambda_{\rm LN}^2}{v^4} \, U \, \bar{m}_\nu^2 \, U^\dagger~.
\eea
We use Greek letters $ \alpha, \beta, \rho, \sigma$ for  the lepton flavor indices, while $i,j$
for the quark flavor indices,
and we neglect  terms with more than two Yukawa insertions.
Moreover, we denote by   $\hat{\alpha}_X$,  $\hat{\beta}_X$, and $\hat{\gamma}_X$  the
numerical coefficients  of $O(1) \times v^2/\Lambda^2$
that multiply the appropriate matrices in flavor space.
For the operators that have a non-vanishing contribution in the $U(3)^5$ limit, we find:
\bea
[\hat{\alpha}_{\varphi l}^{(3)}]^{\alpha \beta}
&=&
\hat{\alpha}_{\varphi l}^{(3)} \, \delta^{\alpha \beta} \ + \
\hat{\beta}_{\varphi l}^{(3)} \,  ( \Delta_{LL}^{(\ell)})^{\alpha \beta} \ + \ \dots \\
V^{im} \, \left[\hat{\alpha}_{\varphi q}^{(3)}\right]^{jm *}
&=&
\hat{\alpha}_{\varphi q}^{(3)} \,  V^{ij}  \ + \
\hat{\beta}_{\varphi q}^{(3)} \,  (V \,  \Delta_{LL}^{(q)})^{ij}  \ + \ \dots \\
V^{im} \, \left[\hat{\alpha}_{l q}^{(3)}\right]^{\alpha \beta mj}
&=&
\hat{\alpha}_{l q}^{(3)} \ \delta^{\alpha \beta} \   V^{ij} +
\hat{\beta}_{l q}^{(3)} \  (\Delta_{LL}^{(\ell)})^{\alpha \beta} \ V^{ij} +
\hat{\gamma}_{l q}^{(3)} \  \delta^{\alpha \beta}\ (V \,  \Delta_{LL}^{(q)})^{ij}
 \ +  \dots \\
\label{eq:alphallnMFV}
\left[\hat{\alpha}_{l l}^{(n)} \right]^{\alpha \beta \rho \sigma} &=&
\hat{\alpha}_{l l}^{(n)} \,   \delta^{\alpha \beta} \ \delta^{\rho \sigma}  +
\hat{\beta}_{l l}^{(n)} \,   \left[ \delta^{\alpha \beta} \  (\Delta_{LL}^{(\ell)})^{\rho \sigma}  +
(\Delta_{LL}^{(\ell)})^{\alpha \beta} \ \delta^{\rho \sigma} \right]  \ + \ \dots \\
\left[\hat{\alpha}_{l e} \right]^{\alpha \beta \rho \sigma} &=&
\hat{\alpha}_{l e} \,  \delta^{\alpha \beta} \delta^{\rho \sigma}   +
\hat{\beta}_{l e} \,  (\Delta_{LL}^{(\ell)})^{\alpha \beta} \, \delta^{\rho \sigma}   \ + \ \dots~.
\eea
For the operators that vanish in the limit of exact $U(3)^5$ symmetry, we find:
\bea
 \left[\hat{\alpha}_{\varphi \varphi} \right]^{ij}
&=&
\hat{\alpha}_{\varphi \varphi} \, \left( \bar{\lambda}_u \, V \, \bar{\lambda}_d \right)^{ij} \ + \ \dots \\
V^{im} \, \left[\hat{\alpha}_{qde}\right]^{\alpha \beta  j m *}
&=&
\hat{\alpha}_{qde} \, \bar{\lambda}_e^{\alpha \beta} \  (V \bar{\lambda}_d)^{ij} \  + \
\hat{\beta}_{qde} \,
 (\bar{\lambda}_e  \, \Delta_{LL}^{(\ell)})^{\alpha \beta} \  (V \bar{\lambda}_d)^{ij} ~ \nonumber \\
 & + &
 \hat{\gamma}_{qde} \,   \bar{\lambda}_e^{\alpha \beta} \  (V \Delta_{LL}^{(q)} \bar{\lambda}_d)^{ij} \  + \
 \dots \\
 \left[\hat{\alpha}_{l q}\right]^{\alpha \beta j i *}
&=&
\hat{\alpha}_{l q} \, \bar{\lambda}_e^{\alpha \beta}\   (\bar{\lambda}_u V)^{ij}  \ + \
\hat{\beta}_{l q} \, (\bar{\lambda}_e \Delta_{LL}^{(\ell)})^{\alpha \beta}\  (\bar{\lambda}_u V)^{ij} \nonumber \\
& + &
\hat{\gamma}_{l q} \, \bar{\lambda}_e^{\alpha \beta}\   (\bar{\lambda}_u V \Delta_{LL}^{(q)})^{ij}
\ +  \ \dots~.
\eea
The coefficient of the tensor operator, $[\alpha^{(t)}_{l q}]$ has an expansion similar to
the one of $[\alpha_{l q}]$.

Except for the top quark, the Yukawa insertions typically involve a large suppression factor, as $\bar{\lambda}_i  = m_i/v$.
In the case  of SM extensions containing two Higgs doublets, this scaling can be modified if there is a hierarchy
between the vacuum expectation values $v_{u}, v_{d}$  of the Higgs fields giving mass to the up- or down-type quarks, respectively.
In this case, for large  $\tan \beta \equiv v_u/v_d$ the Yukawa insertions scale as:
\bea
\bar{\lambda}_u &=& \frac{m_u}{v \, \sin \beta} \  \to \  \frac{m_u}{v}  \\
\bar{\lambda}_d &=& \frac{m_d}{v \, \cos \beta} \  \to \  \frac{m_d}{v}  \, \tan \beta      \\
\bar{\lambda}_\ell &=& \frac{m_\ell }{v \, \cos \beta} \  \to \  \frac{m_\ell}{v}  \, \tan \beta
\eea

\section{Phenomenology of $V_{ud}$ and $V_{us}$: overview}
\label{sect:pheno1}

Using the general effective lagrangians
of Eqs.~(\ref{eq:leffmu}) and (\ref{eq:leffq}) for charged current transitions,
one can calculate the deviations from SM predictions in various semileptonic decays.
In principle a rich phenomenology is possible.  
Helicity suppressed leptonic decays of mesons have been recently analyzed in Ref.~\cite{Filipuzzi:2009xr}.
Concerning semileptonic transitions, 
several reviews treat in some detail
$\beta$ decay differential  distributions~\cite{Herczeg:2001vk,Severijns:2006dr}.
Here  we focus on the integrated decay rates,  which give access to the CKM matrix elements
$V_{ud}$ and $V_{us}$: since both  the SM prediction and the experimental
measurements are reaching the sub-percent level, we expect these observables to
provide strong constraints on NP operators.

$V_{ud}$  and $V_{us}$ can be determined with high precision in a number of channels.
The degree of  needed theoretical input varies, depending on which component of the weak
current contributes to the hadronic  matrix element.  Roughly speaking,
one can group the channels leading to $V_{ud,us}$ into three classes:
\begin{itemize}
\item  semileptonic decays in which only the vector component of the weak current contributes.
These are theoretically favorable in the Standard Model because the matrix elements of
the vector current at zero momentum transfer are known in the $SU(2)$ ($SU(3)$) limit of
equal light quark masses: $m_u=m_d\,  (= m_s)$.   Moreover, corrections to the symmetry limit are quadratic in
$m_{s,d} - m_u$~\cite{Behrends:1960nf,Ademollo:1964sr}.
Super-allowed nuclear beta decays ($0^+ \to 0^+$),   pion beta decay  ($\pi^+ \to \pi^0 e^+  \nu_e$), and
$K \to \pi \ell \nu$ decays belong to this class.
The determination of $V_{ud,us}$ from these modes requires theoretical input
on radiative corrections~\cite{Marciano:1985pd,Marciano:2005ec,Cirigliano:2001mk,Cirigliano:2004pv,Cirigliano:2008wn}
and hadronic matrix elements via analytic methods~\cite{Hardy:2008gy}, \cite{Leutwyler:1984je,Bijnens:2003uy,Jamin:2004re,Cirigliano:2005xn},
or lattice QCD methods~\cite{Becirevic:2004ya,Dawson:2006qc,Boyle:2007qe,Lubicz:2009ht}.

\item semileptonic transitions in which both the vector and axial component of the weak current contribute.
Neutron decay ($n \to p e \bar{\nu}$) and hyperon decays ($\Lambda \to p e \bar{\nu}$, ....) belong to this class.
In this case the matrix elements of the axial current have to be determined experimentally~\cite{Cabibbo:2003cu}.

Inclusive $\tau$ lepton decays  $\tau \to h \nu_\tau$ belong to this class (both V and A current contribute), and
 in this case the relevant matrix elements can be calculated theoretically via the Operator Product
Expansion~\cite{Braaten:1991qm,Gamiz:2002nu}.

\item Leptonic transitions in which only the axial component of the weak current contributes.
In this class one finds meson decays such as $\pi (K)  \to \mu \nu$ but also
exclusive $\tau$ decays such as $\tau \to \nu_\tau \pi (K)$.
Experimentally one can determine the products $V_{ud} \cdot F_\pi$ and $V_{us} \cdot F_K$.
With the advent of precision calculations of $F_K/F_\pi$ in  lattice QCD~\cite{Aubin:2004fs,Beane:2006kx,Follana:2007uv,Blossier:2009bx,Bazavov:2009bb}, this  class of decays provides a useful constraint on the ratio $V_{us}/V_{ud}$~\cite{Marciano:2004uf}.

\end{itemize}

Currently, the determination of $V_{ud}$ is dominated by $0^+ \to 0^+$ super-allowed nuclear beta decays~\cite{Hardy:2008gy},
while  the best determination of $V_{us}$ arises from $K \to \pi \ell \nu$ decays~\cite{Antonelli:2008jg}.
Experimental improvements in neutron decay and $\tau$ decays, as well as in lattice calculations of the decay constants
will allow in the future  competitive determinations from other channels.
In light of this,   we  set out to perform a comprehensive analysis of possible new physics effects
in the extraction of $V_{ud}$ and $V_{us}$.

As outlined in the previous section, we start our analysis by assuming
dominance of the  $U(3)^5$ invariant  operators.
These are not constrained by FCNC and can have a relatively low effective scale $\Lambda$.
In the $U(3)^5$  limit the phenomenology of CC processes  greatly simplifies:
all $V_{ij}$ receive the same  universal  shift (coming from the same short distance structure).
As a consequence, extractions of
$V_{ud,us}$  from  different channels (vector transitions, axial transitions, {\it etc}.)
should agree within errors. Therefore,  in this limit the new physics effects are
entirely captured by the quantity
\be
\label{eq:dckm}
\Delta_{\rm CKM} \equiv  |V_{ud}^{(\rm pheno)}|^2+|V_{us}^{(\rm pheno)}|^2+|V_{ub}^{(\rm pheno)}|^2 \ - \ 1 ~,
\ee
constructed from the $V_{ij}^{\rm (pheno)}$ elements extracted from semileptonic transitions using the standard procedure outlined below.
We now make these points more explicit.

\subsection{Extraction of $V_{ij}$ and contributions to  $\Delta_{\rm CKM}$ in the $U(3)^5$ limit}
If we assume    $U(3)^5$ invariance, only the SM operator survives in the muon decay lagrangian of Eq.~(\ref{eq:leffmu}),  
with~\footnote{We disagree with the result of BW on the sign of $\hat{\alpha}_{ll}^{(3)}$.}
\be
\tilde{v}_L =
4 \, \hat{\alpha}_{\varphi l}^{(3)} - 2 \,  \hat{\alpha}_{ll}^{(3)} ~.
\ee
Therefore, in this case the effect of new physics can be encoded into the following definition of
the leptonic Fermi constant:
\be
G_F^\mu = (G_F)^{(0)}  \, \left(1 +
\tilde{v}_L
 \right)~,
\ee
where $G_F^{(0)} =  g^2/(4 \sqrt{2}  m_W^2)$.
Similarly,   in the  $U(3)^5$  symmetry limit,
only the SM operator survives in the effective langrangian for semileptonic quark decays of Eq.~(\ref{eq:leffq}), with coupling:
\bea
\left[v_L\right]_{\ell \ell ij} & \to &  v_L \equiv  2 \left( \hat{\alpha}_{\varphi l}^{(3)} +
\hat{\alpha}_{\varphi q}^{(3)}- \hat{\alpha}_{l q}^{(3)}\right) ~.
\eea
As in the muon decay, the new physics can be encoded in a (different) shift to the effective semileptonic (SL)  Fermi constant:
\be
G_F^{\rm SL}= (G_F)^{(0)}  \,  \left( 1 +  v_L \right)~.
\ee
The  value of $V_{ij}$ extracted   from semileptonic decays
is affected by this redefinition of
the semileptonic Fermi constant and by the shift in  the  muon Fermi constant $G_F^\mu$, to which one
usually normalizes semileptonic transitions.
In fact one has
\bea
\label{eq:vphenoFB}
V_{ij}^{(\rm pheno)}
&=& V_{ij} \ \frac{G_F^{\rm SL}}{G_F^\mu}  = V_{ij}  \, \left( 1 + v_L - \tilde{v}_L \right)    \nonumber\\
&=& V_{ij} \left[ 1 + 2\, \left(
 \hat{\alpha}_{ll}^{(3)}
- \hat{\alpha}_{l q}^{(3)}
- \hat{\alpha}_{\varphi l}^{(3)} + \hat{\alpha}_{\varphi q}^{(3)}
\right)
 \right]~.
\eea
So in the $U(3)^5$ limit a common shift affects all the $V_{ij}$ (from all channels).
The only way to expose new physics contributions
is to construct universality tests,  in which the absolute normalization of $V_{ij}$ matters.
For light quark  transitions this involves checking that the first row of the CKM matrix is a vector of unit  length
(see definition of  $\Delta_{\rm CKM}$   in  Eq.~(\ref{eq:dckm})).
The new physics  contributions to $\Delta_{\rm CKM}$ involve four operators of our basis and read:
\be
\Delta_{\rm{CKM}}
=             4 \, \left(
\hat{\alpha}_{ll}^{(3)} -\hat{\alpha}_{l q}^{(3)}
- \hat{\alpha}_{\varphi l}^{(3)} + \hat{\alpha}_{\varphi q}^{(3)}
 \right)~.
\label{eq:dckmnp}
\ee
In specific SM extensions, the $\hat{\alpha}_i$ are functions of  the underlying parameters.
Therefore, through the above relation one can work out  the constraints of quark-lepton
universality tests on any weakly coupled SM extension.

\subsection{Beyond $U(3)^5$}

Corrections to the $U(3)^5$ limit can be introduced both within MFV and via generic flavor structures.
In MFV,  as evident from the results of Section~\ref{sect:flavor},  the coefficients parameterizing
deviations from $U(3)^5$ are highly suppressed.
This is true even when one considers  the flavor diagonal
elements of the effective couplings, due to the smallness of the Yukawa eigenvalues and the
hierarchy of the CKM matrix elements.
As a consequence, in MFV we expect the conclusions of the previous subsections to hold.
The various CKM elements $V_{ij}$ receive a common dominant shift plus suppressed channel-dependent
corrections, so that Eq.~(\ref{eq:dckmnp}) remains valid to a good approximation.
In other words, both in the exact $U(3)^5$ limit and
in MFV,   $\Delta_{\rm CKM}$ probes the leading coefficients $\hat{\alpha}_{X}$ of  the four
operators
$O_{\rm CKM} = \{ O_{l l}^{(3)},  O_{l q}^{(3)},  O_{\varphi l}^{(3)},  O_{\varphi q}^{(3)}   \}$.

In a generic non-MFV framework,    the channel-dependent shifts to $V_{ij}$ could be appreciable,
so that $\Delta_{\rm CKM}$ would depend on the channels used to extract $V_{ud,us}$.
Therefore, comparing the values of $V_{us}$ and $V_{ud}$ (or their ratios)  extracted
from different channels gives us a handle on $U(3)^5$ breaking structures
beyond MFV.
We will discuss this  in a  separate publication,  where we will analyze the new physics contributions to
the ratios $V_{ud}^{0^+ \to 0^+}/V_{ud}^{n \to p e \bar{\nu}}$,
$V_{us}^{K \to \pi \ell  \nu}/V_{ud}^{0^+ \to 0^+}$,
$V_{us}^{K \to \mu \nu}/V_{ud}^{\pi \to \mu \nu}$,   and
$(V_{us}//V_{ud})^{\tau \to \nu \, h}$ from both inclusive and exclusive channels.
In summary, we organize our analysis  in two somewhat orthogonal parts, as follows:

\begin{itemize}

\item  In the rest of this work we focus on the phenomenology of  $\Delta_{\rm CKM}$  and
its relation to other precision measurements.
This analysis applies to models of TeV scale physics with
approximate $U(3)^5$ invariance,
in which flavor breaking is suppressed  by  a symmetry principle
(as in MFV)  or  by the hierarchy  $\Lambda_{\rm flavor}  \gg {\rm TeV}$

\item  In a subsequent publication we will explore in detail the constraints arising by
comparing the values  of $V_{us}$ ($V_{ud}$)  extracted from different  channels.
These constraints probe the $U(3)^5$ breaking structures, to which
other precision measurements (especially at high energy) are essentially insensitive.

\end{itemize}

\begin{table}[tb]
\begin{tabular}{|c|c|c|c|}
\hline
 Classification & Standard Notation &  Measurement &Reference\\
 \hline
 Atomic parity &$Q_W(Cs)$&Weak charge in Cs&\cite{Wood:1997zq}\\
 violation ($Q_W$)& $Q_W(Tl)$& Weak charge in Tl&\cite{Vetter:1995vf}\\
 \hline
   DIS&$g_L^2,g_R^2$&$\nu_\mu$-nucleon scattering from NuTeV&\cite{Zeller:2001hh}\\
      & $R^\nu$&$\nu_\mu$-nucleon scattering from CDHS and CHARM&\cite{Blondel:1989ev,Allaby:1986pd}\\
      &$\kappa$&$\nu_\mu$-nucleon scattering from CCFR&\cite{McFarland:1997wx}\\
      &$g_V^{\nu e},g_A^{\nu e}$&$\nu$-$e$ scattering from CHARM II&\cite{Vilain:1994qy}\\

 \hline
   Zline &$\Gamma_Z$&Total $Z$ width&\cite{:2003ih,:2005ema}\\
   (lepton and    &$\sigma_0$ &$e^+e^-$ hadronic cross section at $Z$ pole&\cite{:2003ih,:2005ema}\\
   light quark)      &$R_f^0(f=e,\mu,\tau)$& Ratios of lepton decay rates &\cite{:2003ih,:2005ema}\\
         &$A_{FB}^{0,f}(f=e,\mu,\tau)$ &Forward-backward lepton asymmetries&\cite{:2003ih,:2005ema}\\
 \hline
   pol   &$A_f(f=e,\mu,\tau)$&Polarized lepton asymmetries&\cite{:2003ih,:2005ema}\\
\hline
   bc    &$R_f^0(f=b,c)$& Ratios of hadronic decay rates &\cite{:2003ih,:2005ema}\\
    (heavy quark)     &$A_{FB}^{0,f}(f=b,c)$ &Forward-backward hadronic asymmetries&\cite{:2003ih,:2005ema}\\
	 &$A_f(f=b,c)$&Polarized hadronic asymmetries&\cite{:2003ih,:2005ema}\\
 \hline
 LEPII Fermion      &$\sigma_f(f=q,\mu,\tau)$& Total cross sections for $e^+e^-\rightarrow f\overline f$&\cite{:2003ih,:2005ema}\\
 production   &$A_{FB}^f(f=\mu,\tau)$& Forward-backward asymmetries for $e^+e^-\rightarrow f\overline f$&\cite{:2003ih,:2005ema}\\
\hline
 eOPAL           &$d\sigma_e/d\cos\theta$&Differential cross section for $e^+e^-\rightarrow e^+e^-$&\cite{Abbiendi:2003dh}\\
 \hline
 WL3 &$d\sigma_W/d\cos\theta$& Differential cross section for $e^+e^-\rightarrow W^+W^-$&\cite{Achard:2004zw}\\
 \hline
 MW &$M_W$ & W mass &\cite{:2003ih,:2005ema,Abazov:2003sv}\\
\hline
$Q_{FB}$    &$\sin^2\theta_{eff}^{lept}$&Hadronic charge asymmetry&\cite{:2003ih,:2005ema}\\
\hline
\end{tabular}
\caption{Measurements included in this analysis.
This summary table was taken directly from Table I of  \cite{Han:2004az} and repeated here for convenience.  
We added some details in the classification column as well as additional experimental references.
\label{tab:experiments}}
\end{table}

\section{
$\Delta_{\rm CKM}$ versus  precision electroweak measurements}
\label{sec:InvaraintAnalysis}

In the limit of  approximate $U(3)^5$ invariance, we have shown in  Eq.~(\ref{eq:dckmnp}) that
$\Delta_{\rm CKM}$ constraints a specific  combination of  the coefficients
 $\hat{\alpha}_{l l}^{(3)}, \hat{\alpha}_{l q}^{(3)},
\hat{\alpha}_{\varphi l}^{(3)}, \hat{\alpha}_{\varphi q}^{(3)}$.
Each of these coefficients also contributes to other low- and high-energy  precision electroweak measurements~\cite{Han:2004az},
together with the remaining seventeen operators that make up the $U(3)^5$ invariant sector of our TeV scale
effective lagrangian (see Sect.~\ref{sect:U35invariant}).
Therefore, we can now  address  concrete questions such as:
what is the maximal deviation $|\Delta_{\rm CKM}|$ allowed once all the precision
electroweak constraints have been taken into account?
Which  observables provide the strongest constraints on the  operators contributing to $\Delta_{\rm CKM}$?
How does the inclusion of $\Delta_{\rm CKM}$ affect the fit to precision electroweak measurements?
Should a deviation $\Delta_{\rm CKM} \neq 0$ be established,
in what other precision observables  should we expect
a tension with the SM prediction? At what level?

Our task greatly benefits from the work of Han and Skiba (HS)~\cite{Han:2004az},
who studied the constraints on the same set of twenty-one $U(3)^5$ invariant  operators
via a global fit to precision electroweak data.
We employ a modified version of their publicly available fitting code in what follows.
The analysis utilizes the experimental data summarized in Table \ref{tab:experiments}.
The procedure involves  constructing
the $\chi^2$ function for the observables listed in Table \ref{tab:experiments},
which contains 237 generally correlated terms.
Indicating with $X_{\rm th}^i (\hat{\alpha}_k)$ the theoretical prediction for observable $X^i$ (including SM plus radiative
correction plus first order shift in $\hat{\alpha}_k =  \alpha_k  v^2/\Lambda^2$),
and with $X_{\rm exp}^i$ the experimental value,  the $\chi^2$ reads
\be
\chi^2 (\hat{\alpha}_k) = \sum_{i,j} \,  \Big(X^{i}_{\rm th} (\hat{\alpha}_k) - X^i_{\rm exp} \Big) \ \Big( \sigma^2\Big)^{-1}_{ij} \
  \Big(X^{j}_{\rm th} (\hat{\alpha}_k) - X^j_{\rm exp} \Big)
\ee
where $\sigma_{ij}^2 = \sigma_i \ \rho_{ij} \ \sigma_j$ is expressed in terms of
the combined theoretical and experimental
standard deviation $\sigma_i$ and the correlation matrix  $\rho_{ij}$.
For more details,  we refer to Ref.~\cite{Han:2004az}.
In our numerical analysis we essentially use the code of HS\footnote{
We prefer to quote final results in terms of the dimensionless ratios
$\hat{\alpha}_k =  \alpha_k  v^2/\Lambda^2$ ($v\simeq174$ GeV) instead of $a_k =  1/\Lambda_k^2$ as in HS.
}  and minimally extend it
by including the  $\Delta_{\rm CKM}$ constraint  in  the $\chi^2$ function.
Given the phenomenological input  $V_{ud} = 0.97425(22)$~\cite{Hardy:2008gy},
$V_{us} =0.2252(9)$~\cite{Antonelli:2009ws},
we obtain the constraint
$\Delta_{\rm CKM} = (-1 \pm 6)\times 10^{-4}$~\cite{Antonelli:2009ws}.
$\Delta_{\rm CKM}$  has essentially no correlation with the other precision measurements, due to the small fractional uncertainty in the Fermi constant.

We perform two different analyses, one in which  all operators $O_X$ are allowed to contribute,
and one in which only a single operator at a time  has non vanishing coefficient.
These two regimes represent extreme model scenarios and possess  different characteristics.
In the global analysis, due to the large number of parameters,  cancellations can dilute
the impact of specific observables:  the burden of satisfying
a tight constraint from a given observable can be ``shared'' by several
operators.
On the other hand, within  the single-operator analysis
one may easily find correlations between  different sets of measurements.
We think of the single operator analysis as a survey of a simplified class of models,  in which
only one dominant effective operator is  generated.

\subsection{Global analysis}\label{subsec:GlobalAnalysis}

In order to quantify  the significance of the experimental CKM unitarity constraint,
we first calculate the range of $\Delta_{\rm CKM} (\hat{\alpha}_k)$ allowed by existing
bounds from all the precision electroweak measurements included in Table~\ref{tab:experiments}.
In terms of the best fit values and the  covariance matrix of the $\hat{\alpha}_i$~\cite{Han:2004az}
obtained from the fit to electroweak precision data, we find
\be
 - 9.5 \times 10^{-3}    \ \leq \  \Delta_{\rm CKM}  \  \leq \  0.1  \times 10^{-3} \qquad \qquad
   (90\% \ {\rm C.L.})~,
\ee
to be compared with the direct $90\%$ C.L. bound $|\Delta_{\rm CKM}| \leq  1. \times 10^{-3}$.
The first lesson from this exercise is that electroweak precision data leave
ample room for a sizable non-zero $\Delta_{\rm CKM}$: the direct constraint
is nearly an order of magnitude  stronger than the indirect one!
Therefore,   one should include the $\Delta_{\rm CKM}$ constraint in global fits
to the effective theory parameters.

\begin{figure}
\includegraphics[width=16cm]{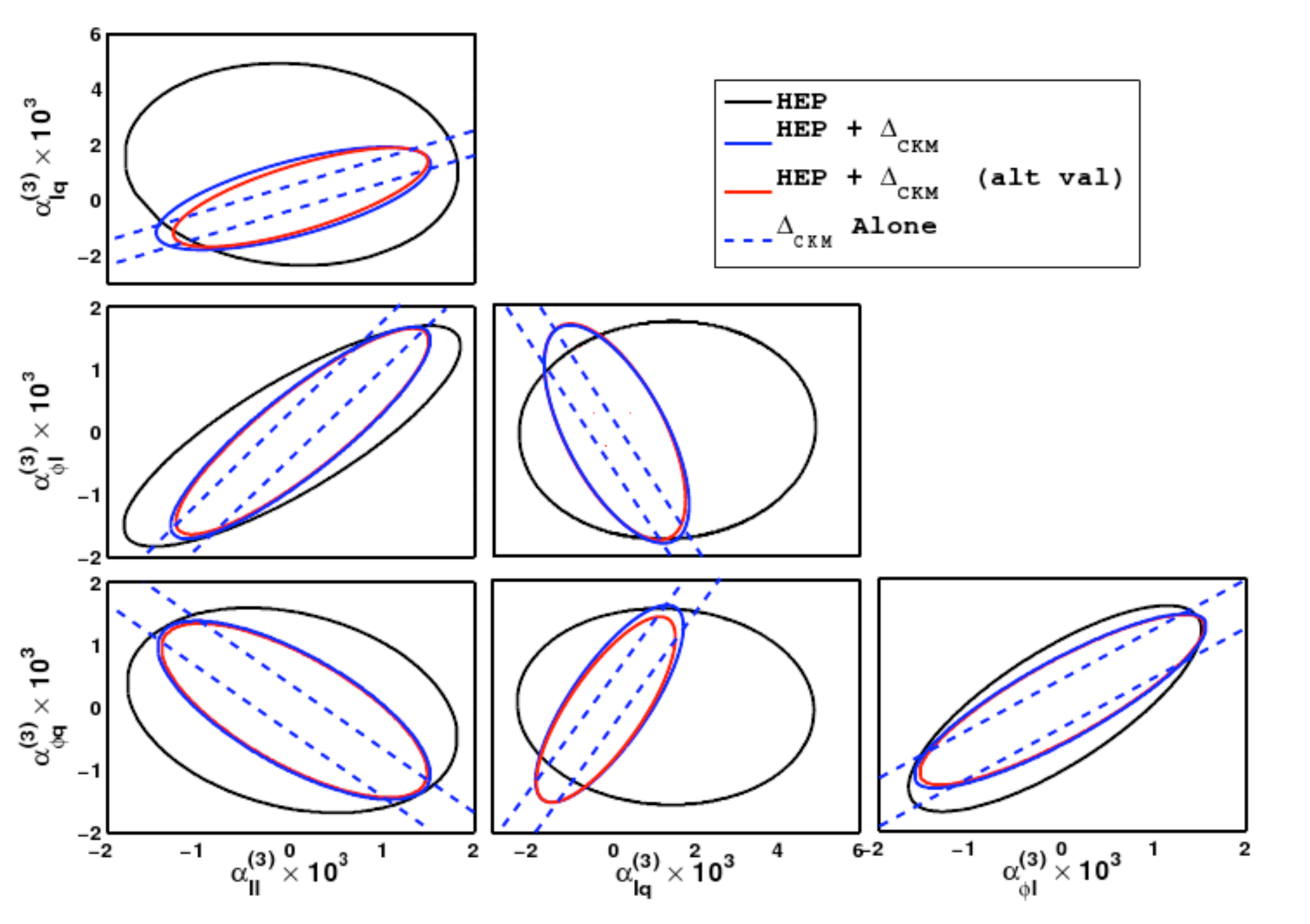}
\caption{$90\%$  allowed regions for the coefficients
$\hat{\alpha}_{ll}^{(3)}, \hat{\alpha}_{l q}^{(3)},
\hat{\alpha}_{\varphi l}^{(3)}, \hat{\alpha}_{\varphi q}^{(3)}$.
These are projections from the $21$ dimensional ellipsoid, obtained from the fitting code.  We include the results for high energy observables alone (HEP, black unbroken curves), high energy data
plus the current $\Delta_{\rm CKM}$ constraint (blue unbroken curve),
high energy data plus the alternative value of $\Delta_{\rm CKM} = -0.0025 \pm 0.0006$ (red unbroken curve) and the bounds from the current $\Delta_{\rm CKM}$ alone (blue dashed curve). \label{fig:DCKMContour} }
\end{figure}

The next question we address is: what is the impact of adding the
$\Delta_{\rm CKM}$ constraint to the global electroweak  fit?
The  chi-squared per degrees of freedom changes only marginally, from 
$\chi^2/d.o.f. =  180.12/215$ to 
$\chi^2/d.o.f. = 173. 74/216$.
We find that essentially the only impact is to modify
the allowed regions for
$\hat{\alpha}_{l l}^{(3)}, \hat{\alpha}_{l q}^{(3)},
\hat{\alpha}_{\varphi l}^{(3)}, \hat{\alpha}_{\varphi q}^{(3)}$.
To illustrate  this, in  Figure~\ref{fig:DCKMContour},
we display the projection of  the  twenty-one dimensional $90\%$  confidence ellipsoid
onto the relevant planes involving
$\hat{\alpha}_{l l}^{(3)}, \hat{\alpha}_{l q}^{(3)},
\hat{\alpha}_{\varphi l}^{(3)}, \hat{\alpha}_{\varphi q}^{(3)}$.
The black curves
represent bounds before the inclusion of the $\Delta_{\rm CKM}$ constraint.
The dashed blue lines outline the allowed regions found by
considering only the effect of current $\Delta_{\rm CKM}$ bounds (Eq.~\ref{eq:dckmnp}):
the regions are unbounded because
large values of any of the  $\hat{\alpha}_i$ may be canceled by a correspondingly large
contribution of other operators.
The situation changes when high energy observables are taken into account, as can be seen from the combined fit solid blue curve.  Despite the relatively weak indirect $\Delta_{\rm CKM}$ constraints
from high energy data, the unbounded parameter directions are cut off at the edge of the allowed black
contour.  In the orthogonal direction, the combined ellipse is shrunk significantly by the strong
$\Delta_{\rm CKM}$ bound.  Thus, the solid blue contour is rotated and contracted with respect to its
parent black region.
As evident from the figure, the main effect of including  $\Delta_{\rm CKM}$
is to strengthen the constraints on  the four-fermion operator $O_{l q}^{(3)}$.

At this stage we may also ask how would this picture change if a significant  deviation from
Cabibbo universality were to be observed.
To answer this question,  we show
in Figure ~\ref{fig:DCKMContour},
the $90 \, \%$ C.L.  allowed regions   (red solid curve)
obtained by assuming a $\sim 4 \, \sigma$
deviation, namely  $\Delta_{\rm CKM} = -0.0025 \pm 0.0006$\footnote{This value has been chosen for illustrative purposes and could be realized
if the central value of $V_{us}$ from $K_{\ell 3}$ decays shifted down to $V_{us}  = 0.2200$,
which is preferred by current analytic estimates of the vector form factor
(see Refs.~\cite{Bijnens:2003uy,Jamin:2004re,Cirigliano:2005xn}).}.
One can see that changing the central value of $\Delta_{\rm CKM}$
has only a minor effect on the allowed regions: the fit is driven by  the comparatively
small $\Delta_{\rm CKM}$ uncertainty,
rather than its central value.
While the fitting  procedure  tends  to minimize the $\chi^2$ contribution from $\Delta_{\rm CKM}$,
this does not generate much tension with the remaining observables,
as other operators can compensate the effect of 
potentially non-vanishint $\hat{\alpha}_i \subset \hat{\alpha}_{\rm CKM}$.

\subsection{Single operator analysis}\label{subsec:IndividualOperators}

To gain a better understanding of the interplay between the $\Delta_{\rm CKM}$ constraint and
other precision measurements, we embark on a single operator analysis.
We assume that a single operator at a time dominates the new physics contribution and set all others to zero.
A similar analysis (not including the CKM constraints) has been performed in~\cite{Barbieri:1999tm}.
We will  only consider the operator set
$O_{\rm CKM} = \{ O_{l l}^{(3)},  O_{l q}^{(3)},  O_{\varphi l}^{(3)},  O_{\varphi q}^{(3)}   \}$
that contributes to $\Delta_{\rm CKM}$,  because for the other operators the analysis
would coincide with that of Ref.~\cite{Barbieri:1999tm}.
In this simplified context we can ask questions about
\begin{itemize}
\item[(i)]  the relative
strength of $\Delta_{\rm CKM}$ versus other precision electroweak measurements
in constraining the non-zero $\hat{\alpha}_i$;
\item[(ii)] the size of correlations  among SM deviations in various observables.
\end{itemize}

\begin{figure}
\includegraphics[scale=0.45]{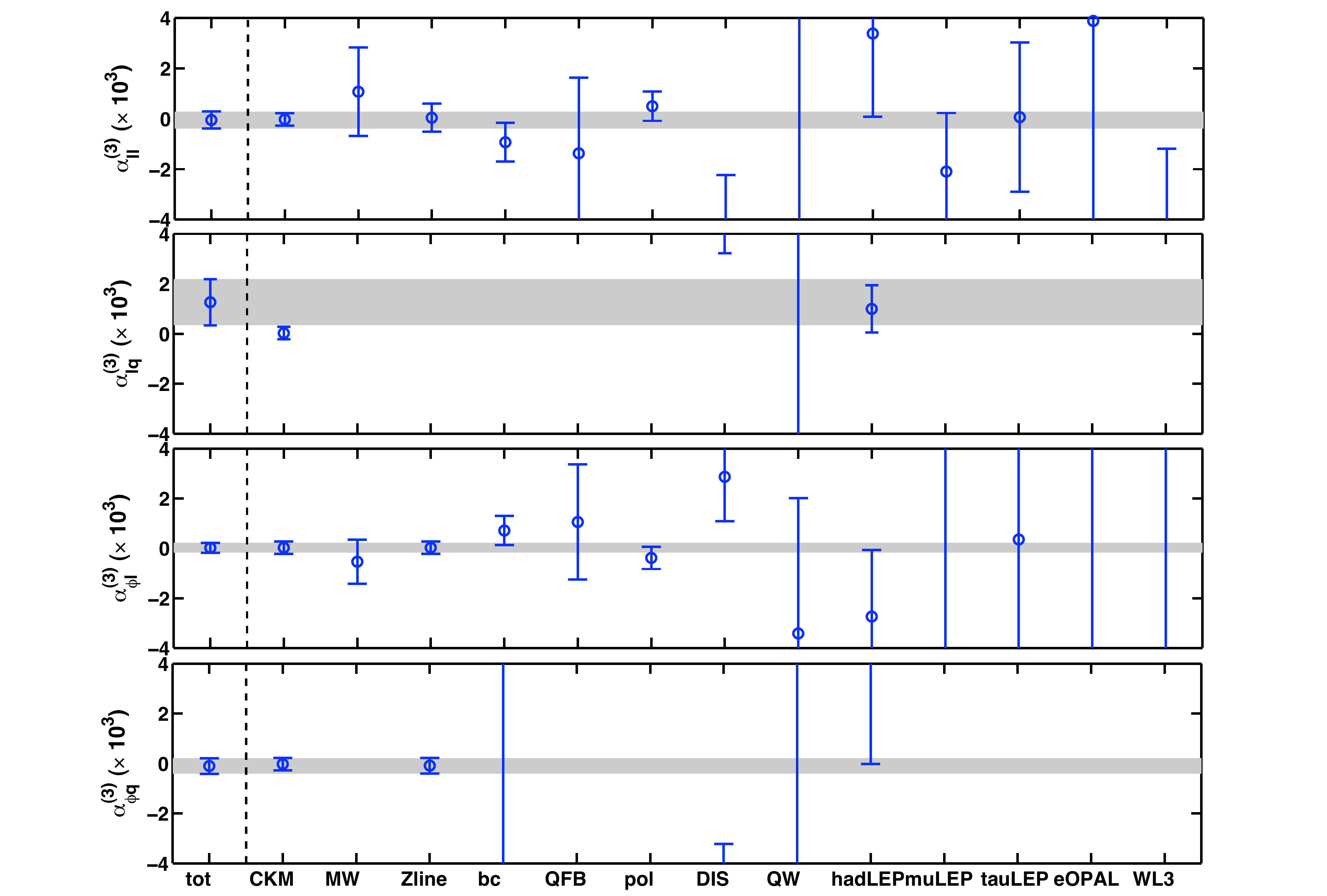}
\caption{The $90 \, \%$ C.L.  allowed regions for the coefficients $\hat{\alpha}_i$
within the single operator analysis.
The first column displays the constraint from all precision observables except $\Delta_{\rm CKM}$.
The second column  displays the constraint coming exclusively from  $\Delta_{\rm CKM}$.
The remaining  columns  display the constraint derived from each subset of
measurements listed in Table~\ref{tab:experiments}. \label{fig:ZoomedIndOpConst}}
\end{figure}

In order to address the first question above,  for each coefficient
$\hat{\alpha}_i \subset \hat{\alpha}_{\rm CKM}$
we derive the  $90 \, \%$ C.L.  allowed intervals implied by:
(a)  the global fit to all  precision electroweak measurements except $\Delta_{\rm CKM}$
(first column in Figure ~\ref{fig:ZoomedIndOpConst}, also denoted by horizontal
gray bands);
(b) the $\Delta_{\rm CKM}$ constraint  via Eq.~(\ref{eq:dckmnp})
(second column in Figure ~\ref{fig:ZoomedIndOpConst});
(c) each  subset of measurements  listed in Table \ref{tab:experiments}
(remaining columns in Figure ~\ref{fig:ZoomedIndOpConst}).
Missing entries in Figure ~(\ref{fig:ZoomedIndOpConst})    signify that
the measurement sets are independent of the selected operator.
The plot nicely illustrates that,  
for the operators $O_i \subset O_{\rm CKM}$,
the direct $\Delta_{\rm CKM}$ measurement 
provides constraints at the same level  (for $\hat{\alpha}_{\varphi l}^{(3)}$) 
or better then the Z pole observables. 
Looking at the size of the constraints,  we can immediately conclude that
the operators $O_{ll}^{(3)}$,
$O_{\varphi l}^{(3)}$,
$O_{\varphi q}^{(3)}$,
are quite  tightly constrained by $Z$ lineshape observables (fourth column in
Figure ~\ref{fig:ZoomedIndOpConst}),
so that very little room is left for  CKM unitarity violations.
On the other hand, the operator $O_{l q}^{(3)}$  is relatively poorly constrained
by electroweak precision data  (LEP2  $e^+ e^- \to q \bar{q}$ cross section provides the best constraint)
and could account for significant deviations of $\Delta_{\rm CKM}$ from zero
(first column of the second panel  from top in Figure ~\ref{fig:ZoomedIndOpConst}). 
In this case,  the direct constraint is by far the tightest.

\begin{figure}
\includegraphics[scale=0.77]{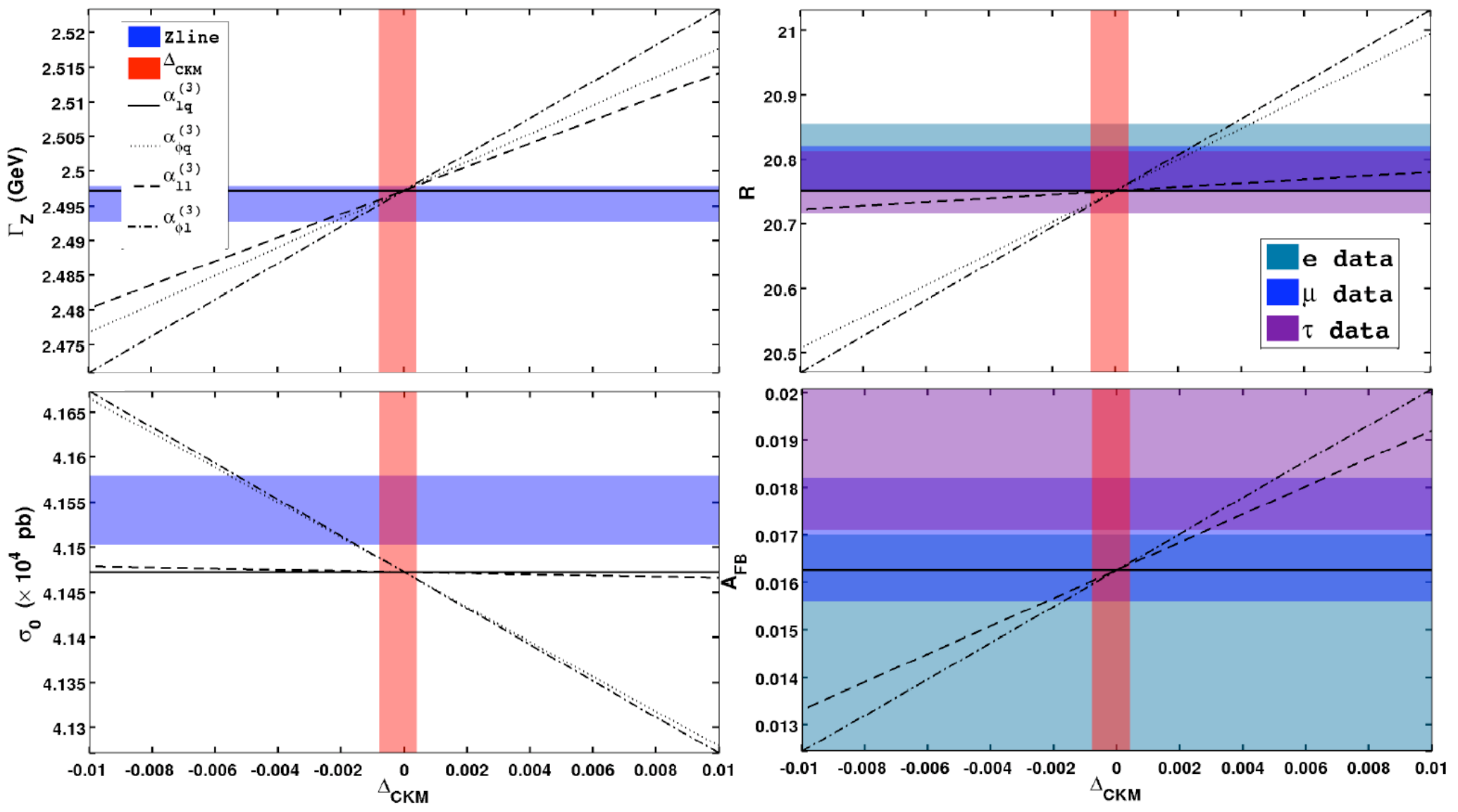}
\caption{Correlation of various $Z$ pole observables with $\Delta_{CKM}$. Operator $O_{lq}^{(3)}$ is not constrained by these measurements.
The    $O_{lq}^{(3)}$  and $O_{\varphi q}^{(3)}$ lines are degenerate in the $A_{FB}$ panel. The $1\sigma$
bands for $\Delta_{\rm CKM}$ and $Z$ pole measurements
are shown in red and blue, respectively.  The right panel
bands  are shaded differently to indicate $e$, $\mu$ and $\tau$ measurements separately.  In the lower left panel $\sigma_0 =   (12\pi\Gamma_{ee}\Gamma_{had})/( M_Z^2\Gamma_Z^2 )$ parameterizes the maximum Z-pole cross-section for $e^+e^- \rightarrow {\rm had}$.\label{fig:ZlineDCKM} }
\end{figure}

\begin{figure}
\includegraphics[scale=0.77]{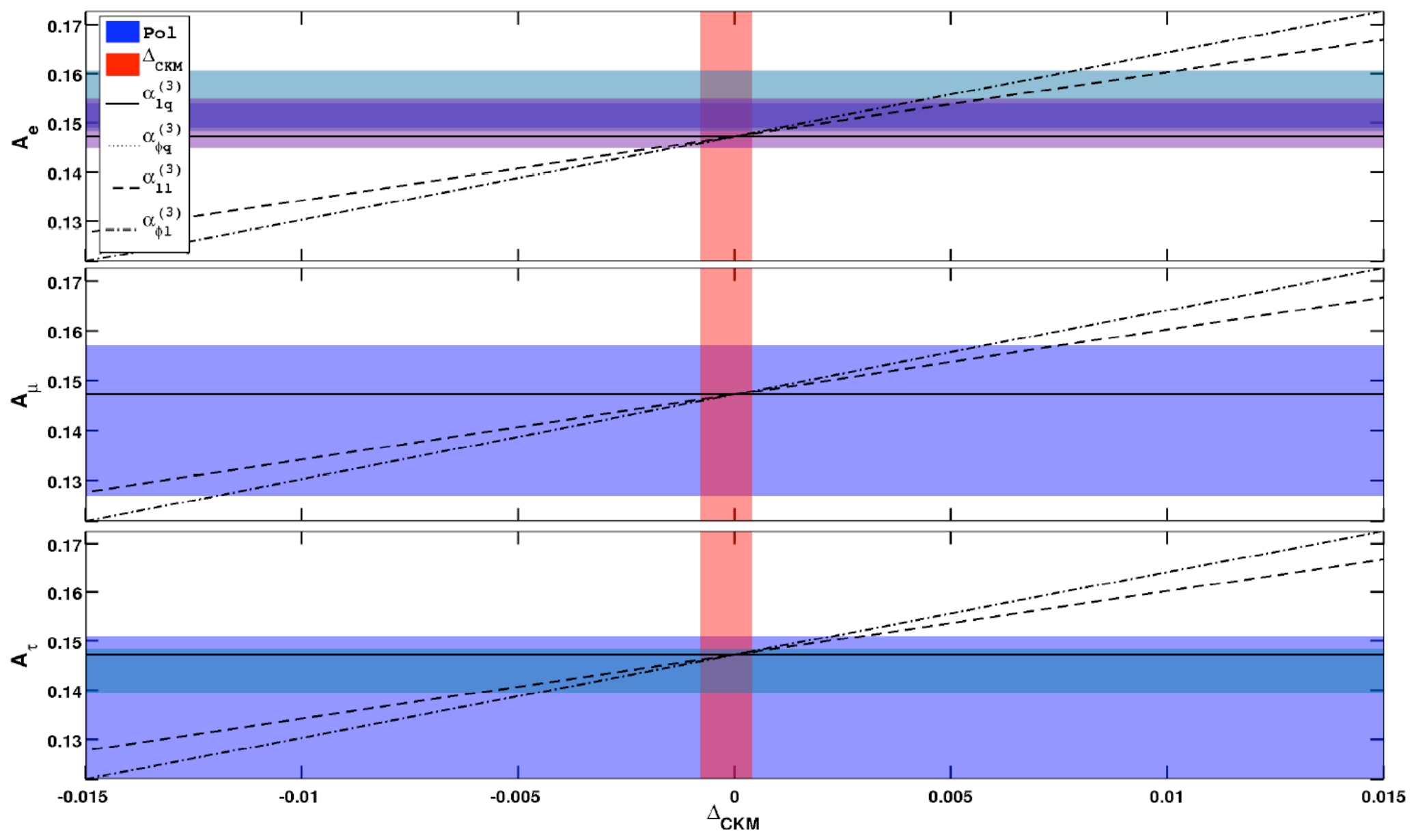}
\caption{Correlation of Z-pole polarized lepton asymmetries
with $\Delta_{\rm CKM}$. Operators $O_{lq}^{(3)}$ and $O_{\varphi q}^{(3)}$ are not constrained by these measurements.  The $1\sigma$ bands for $\Delta_{\rm CKM}$ and lepton asymmetries are shown in red and blue, respectively.
Different blue shading correspond to  different measurements. \label{fig:PolDCKM}}
\end{figure}

Should a non-zero $\Delta_{\rm CKM}$  be observed,
in the single-operator  framework it would be correlated to
deviations from the SM expectation
in other observables as well,
since there is only one parameter in the problem (the coefficient
$\hat{\alpha}_k$ of the dominant operator considered).
We have studied quantitatively the expected correlation between
$\Delta_{\rm CKM}$ and the most sensitive electroweak measurements.
In Figures \ref{fig:ZlineDCKM} and \ref{fig:PolDCKM}
we report the correlation between $\Delta_{\rm CKM}$ and
$Z$ pole observables.
In these figures, each black line (solid or broken) corresponds to
a given single-operator model, in which only one $\hat{\alpha}_{k} \neq 0$.
Each  point on the black line correspond to a particular value of
$\hat{\alpha}_{k}$.
A flat  black line indicate that no correlation exists between the two observables considered.
The red shaded bands indicate the
current 1-$\sigma$  $\Delta_{CKM}$ direct constraint,   while the blue bands
correspond to the  1-$\sigma$  Z-pole  observables.
We use different blue shading to indicate various measurements included in the analysis.
For example, the  forward backward asymmetries ($A_{FB}$) and decay
branching ratios (R) are  shown in different color for each charged lepton flavor.

Figures  \ref{fig:ZlineDCKM} and \ref{fig:PolDCKM}
clearly illustrate how much we can move $\Delta_{\rm CKM}$  from zero
before getting into some tension with Z pole precision measurements.
Moreover, should a given   $\Delta_{\rm CKM} \neq 0$ be measured,
we can immediately read off  in which direction other precision measurement
should move, and by how much, within this class of models.

\begin{figure}
\includegraphics[scale=0.77]{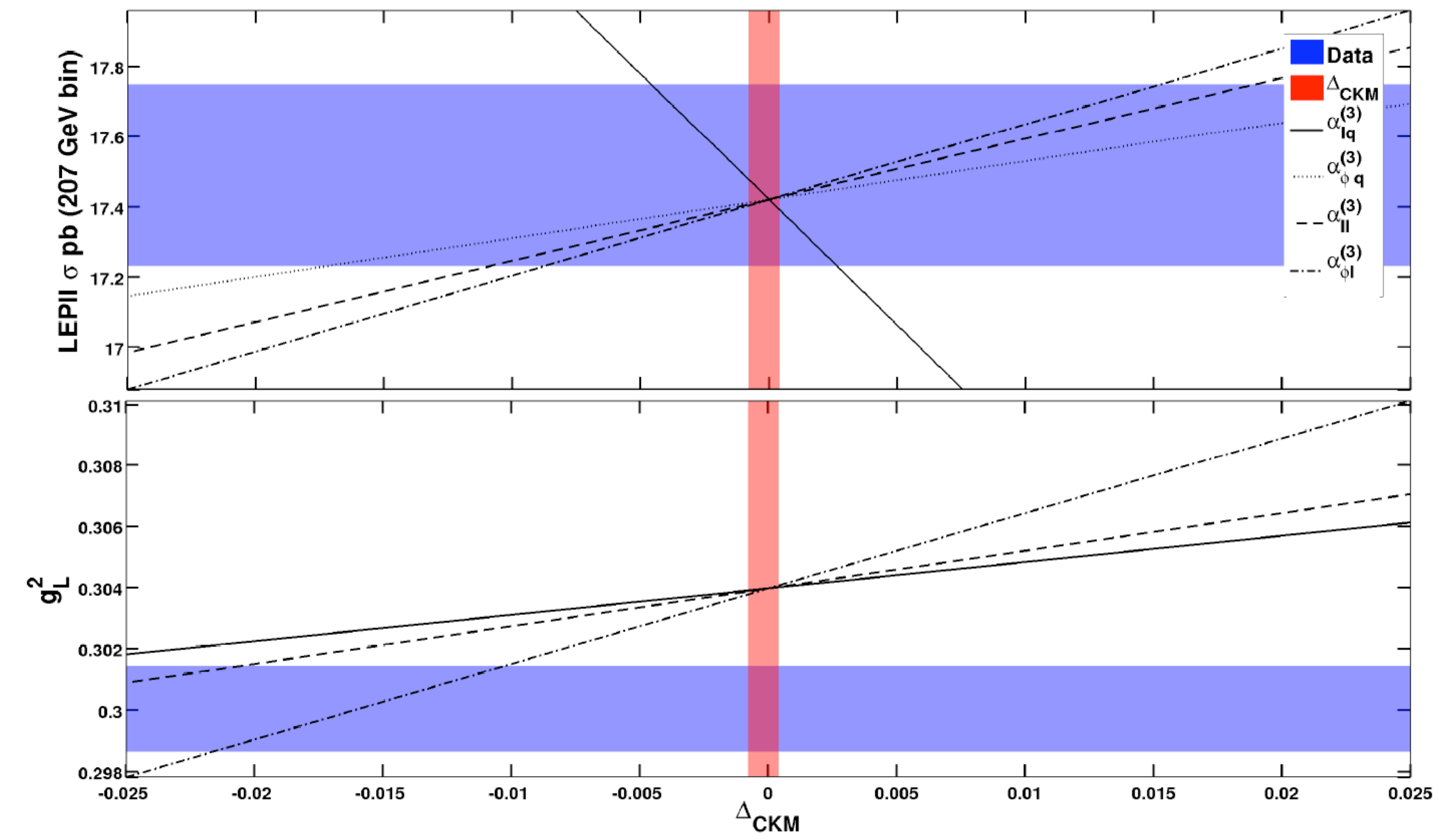}
\caption{Upper panel: correlation between  $\Delta_{\rm CKM}$ and  $\sigma (e^+ e^- \to q \bar{q}) (\sqrt{s}=207 \ {\rm GeV})$.
Lower panel: correlation between $\Delta_{\rm CKM}$ and
the effective neutrino-nucleon couplings $g_{L}^2$ measured by
NuTeV.  The $1\sigma$ bands for  $\Delta_{\rm CKM}$ and the other observable
are shown in red and blue, respectively.\label{fig:NuTeVDCKM}}
\end{figure}

The model in which $O_{l q}^{(3)}$ is the dominant operator is somewhat special,
as  Z-pole observables do not put any constraint.
In this model,  correlations  arise among the following four observables:
$\Delta_{\rm CKM}$,  the LEP2  $e^+ e^- \to q \bar{q}$ cross section,
neutrino DIS  (in particular the NuTeV measurements of the ratios of NC to CC in $\nu_\mu -N $ DIS),
and Atomic Parity Violation, which has only a very weak dependence on $\hat{\alpha}_{l q}^{(3)}$.
The two tightest constraints arise from $\Delta_{\rm CKM}$ and LEP2.
From   the correlation plot in Figure ~\ref{fig:NuTeVDCKM} (upper panel, solid line)
one can see how LEP2 data  in principle leave room for substantial quark-lepton universality violations, up to
$|\Delta_{\rm CKM}| \sim 0.005$ at the 1-$\sigma$ level.
In the lower panel of  Figure ~\ref{fig:NuTeVDCKM}.
we report the correlation plot  between $\Delta_{\rm CKM}$  and the effective neutrino-nucleon
coupling $g_{L}^2$ extracted from NuTeV data.
The striking feature of this plot is that
an explanation of the deviation between the SM prediction and  the NuTeV  measured range  of  $g_{L}^2$
in terms   $O_{l q}^{(3)}$ (solid line)  would require a $\Delta_{\rm CKM}$ at least $16\sigma$ below its current value.

\section{Conclusions}\label{sec:Conclusions}

In this article  we have investigated in a model-independent framework the
impact of  quark-lepton universality tests on probing physics beyond the Standard Model.
We  have identified  a minimal set of twenty-five weak scale effective
operators describing corrections beyond the SM
to precision electroweak measurements and semileptonic decays.
In terms of new physics corrections at the TeV scale, we have derived the low-energy effective lagrangians
describing muon decay and beta decays,  specifying both the
most general flavor structure of the operators as well as the form allowed within Minimal Flavor Violation.

We have performed the phenomenological  analysis assuming nearly  flavor blind
($U(3)^5$ invariant)  new physics interactions.
In this framework  flavor breaking is suppressed  by  a symmetry principle,
such as  the Minimal Flavor Violation hypothesis,  or  by the hierarchy  $\Lambda_{\rm flavor}  \gg {\rm TeV}$.
We have shown that in this limit,  the extraction of $V_{ud}$ and $V_{us}$ from any channel
should give the same result and
the only significant probe of physics beyond the SM involves the quantity  $\Delta_{\rm CKM}  \equiv  |V_{ud}|^2 + |V_{us}|^2 + |V_{ub}|^2 - 1$.
In a subsequent publication we will explore  the constraints arising by
comparing the values  of $V_{us}$ ($V_{ud}$)  extracted from different  channels.
These constraints probe those $U(3)^5$-breaking structures  to which  FCNC and
other precision measurements  are quite  insensitive.

We have shown that in the $U(3)^5$ limit  $\Delta_{\rm CKM}$ receives contributions from four short distance operators,
namely  $O_{\rm CKM} = \{ O_{ll}^{(3)},  O_{l q}^{(3)},  O_{\varphi l}^{(3)},  O_{\varphi q}^{(3)}   \}$,
which also shift SM predictions in other precision observables.
Using the result  of Eq.~\ref{eq:dckmnp}, one can work out the constraints imposed by Cabibbo universality on
any weakly coupled extension of the SM.
Here we have focused on  the model-independent interplay of  $\Delta_{\rm CKM}$  with other precision measurements.
The main conclusions of our analysis are:

\begin{itemize}

\item 
The $\Delta_{\rm CKM}$  constraint  bounds the effective scale of  all  four operators $O_i \subset O_{\rm CKM}$
to be  $\Lambda >  11$ TeV  (90 \% C.L.).
For the operators   $O_{l l}^{(3)},  O_{\varphi l}^{(3)},  O_{\varphi q}^{(3)}$ this constraint is at the same level as 
 the Z-pole measurements.    For the four-fermion operator $O_{l q}^{(3)}$,
$\Delta_{\rm CKM}$  improves existing bounds from LEP2  by one order of magnitude.

\item Another way to state this result is as follows:
should  the central values of $V_{ud}$ and $V_{us}$  move  from the current values~\cite{Antonelli:2008jg},
precision electroweak data would  leave room for sizable deviations from quark-lepton universality
(roughly one order of magnitude above the current direct  constraint).
In a global analysis, the burden of driving a deviation from CKM unitarity  could be  shared by the four operators
$O_i \subset O_{\rm CKM}$.
In a single operator analysis, essentially only the  four-fermion operator $O_{l q}^{(3)}$ could be responsible for $\Delta_{\rm CKM} \neq 0$,
as the others are tightly bound from Z-pole observables.

\end{itemize}
Our conclusions imply that the study of semileptonic  processes and  Cabibbo universality tests
provide constraints  on new physics beyond the SM
that  currently cannot be obtained from other electroweak precision tests and collider measurements.

\appendix

\section{Details on the operator basis}
\label{sec:BWcomments}

In this appendix we discuss how to obtain from the BW operator basis
the minimal subset describing  CP-conserving electroweak precision
observables and beta decays.
We start with a few comments on the BW operator list,
pointing out a few typos and omissions:
\begin{itemize}
\item The four-fermion operator $O^t_{l q} \!=\! (\bar{l}_a\sigma^{\mu\nu}e)\epsilon^{ab}(\bar{q}_b\sigma_{\mu\nu}u)$ must be added to the list
(the $\epsilon$ tensor is used to contract weak SU(2) indices).
\item The operators
$O_{qq}^{(8,1)}, O_{qq}^{(8,3)},  O_{uu}^{(8)}$ and $O_{dd}^{(8)}$
can be eliminated  using the Fierz transformation and the completeness relation of
the Pauli (Gell-Mann) matrices:
$\sum_I\tau^I_{ij}\tau^I_{kl} = - \delta_{ij}\delta_{kl} + 2\delta_{il}\delta_{kj}$;
\item The dagger in the operator (3.55) should be replaced by a T (transpose symbol);
\item The names $O_{uG}$ and $O_{dG}$ have been used twice in BW:
operators (3.34,  3.36) and operators (3.61, 3.63).
\end{itemize}
As a result of  the above  observations, the complete list of dimension six operators involves
seventy-seven operators.

Once the CP-assumption is taken into account, we have
seventy-one
operators in our effective lagrangian\footnote{The six operators removed are $O_X$ with $X=\tilde{G},\tilde{W},\varphi\tilde{G},\varphi\tilde{W},\varphi\tilde{B},\tilde{W}B$.}.
Moreover, we will not take into account the thirteen
operators that involve only quark and gluon fields\footnote{$O_X$ with $X=G,qq^{(1)},qq^{(8)},uu^{(1)},dd^{(1)},qq^{(1,1)},qq^{(1,3)},ud^{(1)},ud^{(8)},qu^{(1)},qu^{(8)},$ $qd^{(1)},qd^{(8)}$.}, because they will not appear in our observables (precision EW measurements and semileptonic decays) at the level we are working.
Further  operators that do not contribute to our observables
are $O_{qG},O_{uG},O_{dG}$.

Since we are not considering processes involving the Higgs boson as an external particle,
we can remove more operators from our list:
 $O_\varphi, O_{\partial\varphi}$ (they only involve scalar fields), and
 seven more operators\footnote{$O_X$ with $X=\varphi W, \varphi B, \varphi^{(1)}, \varphi G, e\varphi, u\varphi, d\varphi$} whose effect  can be absorbed in a redefinition of the SM parameters $g$, $g'$, $g_s$,  $v$ and the Yukawa couplings.
In this way we end up with forty-six
operators that can produce a linear correction to the SM-prediction of our observables. But a more detailed analysis
of  this list  shows that twenty-one of them either do not produce linear corrections (because the interference with the SM vanishes) or produce
effects suppressed by an additional factor (for example, low energy  four-quark
operators of dimension seven).

Finally we have the twenty-five
operators listed in the text:  twenty-one  of them are
invariant under the flavor symmetry $U(3)^5$ and contribute without suppression
to the precision EW measurements~\cite{Han:2004az}.
The remaining four operators are non-invariant under $U(3)^5$.

\acknowledgments{
We thank the Institute for Nuclear Theory at the University of
Washington and the Aspen Center for Physics for their hospitality
during the  completion of this  work.
We thank Antonio Pich and Jorge Portol\'es for useful comments. 
M.G.-A. thanks A. Filipuzzi for useful comments and discussions and the LANL T-2 Group for its hospitality and partial support. This work has been  
supported in part by the EU RTN network FLAVIAnet [Contract No.  
MRTN-CT-2006-035482], by MICINN, Spain [Grants FPU No. AP20050910,  
FPA2007-60323 and Consolider-Ingenio 2010 Programme CSD2007-00042  
-CPAN-] (M. G.-A.).
 This work was  performed under the auspices of the National Nuclear Security Administration of the U.S. Department of Energy at Los Alamos National Laboratory under Contract No. DE-AC52-06NA25396,
 and was supported in part by the LANL LDRD program.

\bibliographystyle{apsrev}

\bibliography{ReferencesPublic}

\begin{thebibliography}{64}
\expandafter\ifx\csname natexlab\endcsname\relax\def\natexlab#1{#1}\fi
\expandafter\ifx\csname bibnamefont\endcsname\relax
  \def\bibnamefont#1{#1}\fi
\expandafter\ifx\csname bibfnamefont\endcsname\relax
  \def\bibfnamefont#1{#1}\fi
\expandafter\ifx\csname citenamefont\endcsname\relax
  \def\citenamefont#1{#1}\fi
\expandafter\ifx\csname url\endcsname\relax
  \def\url#1{\texttt{#1}}\fi
\expandafter\ifx\csname urlprefix\endcsname\relax\def\urlprefix{URL }\fi
\providecommand{\bibinfo}[2]{#2}
\providecommand{\eprint}[2][]{\url{#2}}

\bibitem[{\citenamefont{Amsler et~al.}(2008)}]{Amsler:2008zzb}
\bibinfo{author}{\bibfnamefont{C.}~\bibnamefont{Amsler}} \bibnamefont{et~al.}
  (\bibinfo{collaboration}{Particle Data Group}), \bibinfo{journal}{Phys.
  Lett.} \textbf{\bibinfo{volume}{B667}}, \bibinfo{pages}{1}
  (\bibinfo{year}{2008}).

\bibitem[{\citenamefont{Marciano}(2008)}]{Marciano:2007zz}
\bibinfo{author}{\bibfnamefont{W.~J.} \bibnamefont{Marciano}},
  \bibinfo{journal}{PoS} \textbf{\bibinfo{volume}{KAON}}, \bibinfo{pages}{003}
  (\bibinfo{year}{2008}).

\bibitem[{\citenamefont{Antonelli et~al.}(2008)}]{Antonelli:2008jg}
\bibinfo{author}{\bibfnamefont{M.}~\bibnamefont{Antonelli}}
  \bibnamefont{et~al.} (\bibinfo{collaboration}{FlaviaNet Working Group on Kaon
  Decays}) (\bibinfo{year}{2008}), \eprint{0801.1817}.

\bibitem[{\citenamefont{Cabibbo}(1963)}]{Cabibbo:1963yz}
\bibinfo{author}{\bibfnamefont{N.}~\bibnamefont{Cabibbo}},
  \bibinfo{journal}{Phys. Rev. Lett.} \textbf{\bibinfo{volume}{10}},
  \bibinfo{pages}{531} (\bibinfo{year}{1963}).

\bibitem[{\citenamefont{Kobayashi and Maskawa}(1973)}]{Kobayashi:1973fv}
\bibinfo{author}{\bibfnamefont{M.}~\bibnamefont{Kobayashi}} \bibnamefont{and}
  \bibinfo{author}{\bibfnamefont{T.}~\bibnamefont{Maskawa}},
  \bibinfo{journal}{Prog. Theor. Phys.} \textbf{\bibinfo{volume}{49}},
  \bibinfo{pages}{652} (\bibinfo{year}{1973}).

\bibitem[{\citenamefont{Barbieri et~al.}(1985)\citenamefont{Barbieri, Bouchiat,
  Georges, and Le~Doussal}}]{Barbieri:1985ff}
\bibinfo{author}{\bibfnamefont{R.}~\bibnamefont{Barbieri}},
  \bibinfo{author}{\bibfnamefont{C.}~\bibnamefont{Bouchiat}},
  \bibinfo{author}{\bibfnamefont{A.}~\bibnamefont{Georges}}, \bibnamefont{and}
  \bibinfo{author}{\bibfnamefont{P.}~\bibnamefont{Le~Doussal}},
  \bibinfo{journal}{Phys. Lett.} \textbf{\bibinfo{volume}{B156}},
  \bibinfo{pages}{348} (\bibinfo{year}{1985}).

\bibitem[{\citenamefont{Marciano and Sirlin}(1987)}]{Marciano:1987ja}
\bibinfo{author}{\bibfnamefont{W.~J.} \bibnamefont{Marciano}} \bibnamefont{and}
  \bibinfo{author}{\bibfnamefont{A.}~\bibnamefont{Sirlin}},
  \bibinfo{journal}{Phys. Rev.} \textbf{\bibinfo{volume}{D35}},
  \bibinfo{pages}{1672} (\bibinfo{year}{1987}).

\bibitem[{\citenamefont{Hagiwara et~al.}(1995)\citenamefont{Hagiwara,
  Matsumoto, and Yamada}}]{Hagiwara:1995fx}
\bibinfo{author}{\bibfnamefont{K.}~\bibnamefont{Hagiwara}},
  \bibinfo{author}{\bibfnamefont{S.}~\bibnamefont{Matsumoto}},
  \bibnamefont{and} \bibinfo{author}{\bibfnamefont{Y.}~\bibnamefont{Yamada}},
  \bibinfo{journal}{Phys. Rev. Lett.} \textbf{\bibinfo{volume}{75}},
  \bibinfo{pages}{3605} (\bibinfo{year}{1995}), \eprint{hep-ph/9507419}.

\bibitem[{\citenamefont{Kurylov and Ramsey-Musolf}(2002)}]{Kurylov:2001zx}
\bibinfo{author}{\bibfnamefont{A.}~\bibnamefont{Kurylov}} \bibnamefont{and}
  \bibinfo{author}{\bibfnamefont{M.~J.} \bibnamefont{Ramsey-Musolf}},
  \bibinfo{journal}{Phys. Rev. Lett.} \textbf{\bibinfo{volume}{88}},
  \bibinfo{pages}{071804} (\bibinfo{year}{2002}), \eprint{hep-ph/0109222}.

\bibitem[{\citenamefont{Buchmuller and Wyler}(1986)}]{Buchmuller:1985jz}
\bibinfo{author}{\bibfnamefont{W.}~\bibnamefont{Buchmuller}} \bibnamefont{and}
  \bibinfo{author}{\bibfnamefont{D.}~\bibnamefont{Wyler}},
  \bibinfo{journal}{Nucl. Phys.} \textbf{\bibinfo{volume}{B268}},
  \bibinfo{pages}{621} (\bibinfo{year}{1986}).

\bibitem[{\citenamefont{Appelquist and Wu}(1993)}]{Appelquist:1993ka}
\bibinfo{author}{\bibfnamefont{T.}~\bibnamefont{Appelquist}} \bibnamefont{and}
  \bibinfo{author}{\bibfnamefont{G.-H.} \bibnamefont{Wu}},
  \bibinfo{journal}{Phys. Rev.} \textbf{\bibinfo{volume}{D48}},
  \bibinfo{pages}{3235} (\bibinfo{year}{1993}), \eprint{hep-ph/9304240}.

\bibitem[{\citenamefont{Longhitano}(1980)}]{Longhitano:1980iz}
\bibinfo{author}{\bibfnamefont{A.~C.} \bibnamefont{Longhitano}},
  \bibinfo{journal}{Phys. Rev.} \textbf{\bibinfo{volume}{D22}},
  \bibinfo{pages}{1166} (\bibinfo{year}{1980}).

\bibitem[{\citenamefont{Feruglio}(1993)}]{Feruglio:1992wf}
\bibinfo{author}{\bibfnamefont{F.}~\bibnamefont{Feruglio}},
  \bibinfo{journal}{Int. J. Mod. Phys.} \textbf{\bibinfo{volume}{A8}},
  \bibinfo{pages}{4937} (\bibinfo{year}{1993}), \eprint{hep-ph/9301281}.

\bibitem[{\citenamefont{Wudka}(1994)}]{Wudka:1994ny}
\bibinfo{author}{\bibfnamefont{J.}~\bibnamefont{Wudka}}, \bibinfo{journal}{Int.
  J. Mod. Phys.} \textbf{\bibinfo{volume}{A9}}, \bibinfo{pages}{2301}
  (\bibinfo{year}{1994}), \eprint{hep-ph/9406205}.

\bibitem[{\citenamefont{de~Gouvea and Jenkins}(2008)}]{deGouvea:2007xp}
\bibinfo{author}{\bibfnamefont{A.}~\bibnamefont{de~Gouvea}} \bibnamefont{and}
  \bibinfo{author}{\bibfnamefont{J.}~\bibnamefont{Jenkins}},
  \bibinfo{journal}{Phys. Rev.} \textbf{\bibinfo{volume}{D77}},
  \bibinfo{pages}{013008} (\bibinfo{year}{2008}), \eprint{0708.1344}.

\bibitem[{\citenamefont{Han and Skiba}(2005)}]{Han:2004az}
\bibinfo{author}{\bibfnamefont{Z.}~\bibnamefont{Han}} \bibnamefont{and}
  \bibinfo{author}{\bibfnamefont{W.}~\bibnamefont{Skiba}},
  \bibinfo{journal}{Phys. Rev.} \textbf{\bibinfo{volume}{D71}},
  \bibinfo{pages}{075009} (\bibinfo{year}{2005}), \eprint{hep-ph/0412166}.

\bibitem[{\citenamefont{Chivukula and Georgi}(1987)}]{Chivukula:1987py}
\bibinfo{author}{\bibfnamefont{R.~S.} \bibnamefont{Chivukula}}
  \bibnamefont{and} \bibinfo{author}{\bibfnamefont{H.}~\bibnamefont{Georgi}},
  \bibinfo{journal}{Phys. Lett.} \textbf{\bibinfo{volume}{B188}},
  \bibinfo{pages}{99} (\bibinfo{year}{1987}).

\bibitem[{\citenamefont{Hall and Randall}(1990)}]{Hall:1990ac}
\bibinfo{author}{\bibfnamefont{L.~J.} \bibnamefont{Hall}} \bibnamefont{and}
  \bibinfo{author}{\bibfnamefont{L.}~\bibnamefont{Randall}},
  \bibinfo{journal}{Phys. Rev. Lett.} \textbf{\bibinfo{volume}{65}},
  \bibinfo{pages}{2939} (\bibinfo{year}{1990}).

\bibitem[{\citenamefont{Buras et~al.}(2001)\citenamefont{Buras, Gambino,
  Gorbahn, Jager, and Silvestrini}}]{Buras:2000dm}
\bibinfo{author}{\bibfnamefont{A.~J.} \bibnamefont{Buras}},
  \bibinfo{author}{\bibfnamefont{P.}~\bibnamefont{Gambino}},
  \bibinfo{author}{\bibfnamefont{M.}~\bibnamefont{Gorbahn}},
  \bibinfo{author}{\bibfnamefont{S.}~\bibnamefont{Jager}}, \bibnamefont{and}
  \bibinfo{author}{\bibfnamefont{L.}~\bibnamefont{Silvestrini}},
  \bibinfo{journal}{Phys. Lett.} \textbf{\bibinfo{volume}{B500}},
  \bibinfo{pages}{161} (\bibinfo{year}{2001}), \eprint{hep-ph/0007085}.

\bibitem[{\citenamefont{D'Ambrosio et~al.}(2002)\citenamefont{D'Ambrosio,
  Giudice, Isidori, and Strumia}}]{D'Ambrosio:2002ex}
\bibinfo{author}{\bibfnamefont{G.}~\bibnamefont{D'Ambrosio}},
  \bibinfo{author}{\bibfnamefont{G.~F.} \bibnamefont{Giudice}},
  \bibinfo{author}{\bibfnamefont{G.}~\bibnamefont{Isidori}}, \bibnamefont{and}
  \bibinfo{author}{\bibfnamefont{A.}~\bibnamefont{Strumia}},
  \bibinfo{journal}{Nucl. Phys.} \textbf{\bibinfo{volume}{B645}},
  \bibinfo{pages}{155} (\bibinfo{year}{2002}), \eprint{hep-ph/0207036}.

\bibitem[{\citenamefont{Cirigliano
  et~al.}(2005{\natexlab{a}})\citenamefont{Cirigliano, Grinstein, Isidori, and
  Wise}}]{Cirigliano:2005ck}
\bibinfo{author}{\bibfnamefont{V.}~\bibnamefont{Cirigliano}},
  \bibinfo{author}{\bibfnamefont{B.}~\bibnamefont{Grinstein}},
  \bibinfo{author}{\bibfnamefont{G.}~\bibnamefont{Isidori}}, \bibnamefont{and}
  \bibinfo{author}{\bibfnamefont{M.~B.} \bibnamefont{Wise}},
  \bibinfo{journal}{Nucl. Phys.} \textbf{\bibinfo{volume}{B728}},
  \bibinfo{pages}{121} (\bibinfo{year}{2005}{\natexlab{a}}),
  \eprint{hep-ph/0507001}.

\bibitem[{\citenamefont{Maki et~al.}(1962)\citenamefont{Maki, Nakagawa, and
  Sakata}}]{Maki:1962mu}
\bibinfo{author}{\bibfnamefont{Z.}~\bibnamefont{Maki}},
  \bibinfo{author}{\bibfnamefont{M.}~\bibnamefont{Nakagawa}}, \bibnamefont{and}
  \bibinfo{author}{\bibfnamefont{S.}~\bibnamefont{Sakata}},
  \bibinfo{journal}{Prog. Theor. Phys.} \textbf{\bibinfo{volume}{28}},
  \bibinfo{pages}{870} (\bibinfo{year}{1962}).

\bibitem[{\citenamefont{Filipuzzi and Isidori}(2009)}]{Filipuzzi:2009xr}
\bibinfo{author}{\bibfnamefont{A.}~\bibnamefont{Filipuzzi}} \bibnamefont{and}
  \bibinfo{author}{\bibfnamefont{G.}~\bibnamefont{Isidori}}
  (\bibinfo{year}{2009}), \eprint{0906.3024}.

\bibitem[{\citenamefont{Herczeg}(2001)}]{Herczeg:2001vk}
\bibinfo{author}{\bibfnamefont{P.}~\bibnamefont{Herczeg}},
  \bibinfo{journal}{Prog. Part. Nucl. Phys.} \textbf{\bibinfo{volume}{46}},
  \bibinfo{pages}{413} (\bibinfo{year}{2001}).

\bibitem[{\citenamefont{Severijns et~al.}(2006)\citenamefont{Severijns, Beck,
  and Naviliat-Cuncic}}]{Severijns:2006dr}
\bibinfo{author}{\bibfnamefont{N.}~\bibnamefont{Severijns}},
  \bibinfo{author}{\bibfnamefont{M.}~\bibnamefont{Beck}}, \bibnamefont{and}
  \bibinfo{author}{\bibfnamefont{O.}~\bibnamefont{Naviliat-Cuncic}},
  \bibinfo{journal}{Rev. Mod. Phys.} \textbf{\bibinfo{volume}{78}},
  \bibinfo{pages}{991} (\bibinfo{year}{2006}), \eprint{nucl-ex/0605029}.

\bibitem[{\citenamefont{Behrends and Sirlin}(1960)}]{Behrends:1960nf}
\bibinfo{author}{\bibfnamefont{R.~E.} \bibnamefont{Behrends}} \bibnamefont{and}
  \bibinfo{author}{\bibfnamefont{A.}~\bibnamefont{Sirlin}},
  \bibinfo{journal}{Phys. Rev. Lett.} \textbf{\bibinfo{volume}{4}},
  \bibinfo{pages}{186} (\bibinfo{year}{1960}).

\bibitem[{\citenamefont{Ademollo and Gatto}(1964)}]{Ademollo:1964sr}
\bibinfo{author}{\bibfnamefont{M.}~\bibnamefont{Ademollo}} \bibnamefont{and}
  \bibinfo{author}{\bibfnamefont{R.}~\bibnamefont{Gatto}},
  \bibinfo{journal}{Phys. Rev. Lett.} \textbf{\bibinfo{volume}{13}},
  \bibinfo{pages}{264} (\bibinfo{year}{1964}).

\bibitem[{\citenamefont{Marciano and Sirlin}(1986)}]{Marciano:1985pd}
\bibinfo{author}{\bibfnamefont{W.~J.} \bibnamefont{Marciano}} \bibnamefont{and}
  \bibinfo{author}{\bibfnamefont{A.}~\bibnamefont{Sirlin}},
  \bibinfo{journal}{Phys. Rev. Lett.} \textbf{\bibinfo{volume}{56}},
  \bibinfo{pages}{22} (\bibinfo{year}{1986}).

\bibitem[{\citenamefont{Marciano and Sirlin}(2006)}]{Marciano:2005ec}
\bibinfo{author}{\bibfnamefont{W.~J.} \bibnamefont{Marciano}} \bibnamefont{and}
  \bibinfo{author}{\bibfnamefont{A.}~\bibnamefont{Sirlin}},
  \bibinfo{journal}{Phys. Rev. Lett.} \textbf{\bibinfo{volume}{96}},
  \bibinfo{pages}{032002} (\bibinfo{year}{2006}), \eprint{hep-ph/0510099}.

\bibitem[{\citenamefont{Cirigliano et~al.}(2002)\citenamefont{Cirigliano,
  Knecht, Neufeld, Rupertsberger, and Talavera}}]{Cirigliano:2001mk}
\bibinfo{author}{\bibfnamefont{V.}~\bibnamefont{Cirigliano}},
  \bibinfo{author}{\bibfnamefont{M.}~\bibnamefont{Knecht}},
  \bibinfo{author}{\bibfnamefont{H.}~\bibnamefont{Neufeld}},
  \bibinfo{author}{\bibfnamefont{H.}~\bibnamefont{Rupertsberger}},
  \bibnamefont{and} \bibinfo{author}{\bibfnamefont{P.}~\bibnamefont{Talavera}},
  \bibinfo{journal}{Eur. Phys. J.} \textbf{\bibinfo{volume}{C23}},
  \bibinfo{pages}{121} (\bibinfo{year}{2002}), \eprint{hep-ph/0110153}.

\bibitem[{\citenamefont{Cirigliano et~al.}(2004)\citenamefont{Cirigliano,
  Neufeld, and Pichl}}]{Cirigliano:2004pv}
\bibinfo{author}{\bibfnamefont{V.}~\bibnamefont{Cirigliano}},
  \bibinfo{author}{\bibfnamefont{H.}~\bibnamefont{Neufeld}}, \bibnamefont{and}
  \bibinfo{author}{\bibfnamefont{H.}~\bibnamefont{Pichl}},
  \bibinfo{journal}{Eur. Phys. J.} \textbf{\bibinfo{volume}{C35}},
  \bibinfo{pages}{53} (\bibinfo{year}{2004}), \eprint{hep-ph/0401173}.

\bibitem[{\citenamefont{Cirigliano et~al.}(2008)\citenamefont{Cirigliano,
  Giannotti, and Neufeld}}]{Cirigliano:2008wn}
\bibinfo{author}{\bibfnamefont{V.}~\bibnamefont{Cirigliano}},
  \bibinfo{author}{\bibfnamefont{M.}~\bibnamefont{Giannotti}},
  \bibnamefont{and} \bibinfo{author}{\bibfnamefont{H.}~\bibnamefont{Neufeld}},
  \bibinfo{journal}{JHEP} \textbf{\bibinfo{volume}{11}}, \bibinfo{pages}{006}
  (\bibinfo{year}{2008}), \eprint{0807.4507}.

\bibitem[{\citenamefont{Hardy and Towner}(2008)}]{Hardy:2008gy}
\bibinfo{author}{\bibfnamefont{J.~C.} \bibnamefont{Hardy}} \bibnamefont{and}
  \bibinfo{author}{\bibfnamefont{I.~S.} \bibnamefont{Towner}}
  (\bibinfo{year}{2008}), \eprint{0812.1202}.

\bibitem[{\citenamefont{Leutwyler and Roos}(1984)}]{Leutwyler:1984je}
\bibinfo{author}{\bibfnamefont{H.}~\bibnamefont{Leutwyler}} \bibnamefont{and}
  \bibinfo{author}{\bibfnamefont{M.}~\bibnamefont{Roos}}, \bibinfo{journal}{Z.
  Phys.} \textbf{\bibinfo{volume}{C25}}, \bibinfo{pages}{91}
  (\bibinfo{year}{1984}).

\bibitem[{\citenamefont{Bijnens and Talavera}(2003)}]{Bijnens:2003uy}
\bibinfo{author}{\bibfnamefont{J.}~\bibnamefont{Bijnens}} \bibnamefont{and}
  \bibinfo{author}{\bibfnamefont{P.}~\bibnamefont{Talavera}},
  \bibinfo{journal}{Nucl. Phys.} \textbf{\bibinfo{volume}{B669}},
  \bibinfo{pages}{341} (\bibinfo{year}{2003}), \eprint{hep-ph/0303103}.

\bibitem[{\citenamefont{Jamin et~al.}(2004)\citenamefont{Jamin, Oller, and
  Pich}}]{Jamin:2004re}
\bibinfo{author}{\bibfnamefont{M.}~\bibnamefont{Jamin}},
  \bibinfo{author}{\bibfnamefont{J.~A.} \bibnamefont{Oller}}, \bibnamefont{and}
  \bibinfo{author}{\bibfnamefont{A.}~\bibnamefont{Pich}},
  \bibinfo{journal}{JHEP} \textbf{\bibinfo{volume}{02}}, \bibinfo{pages}{047}
  (\bibinfo{year}{2004}), \eprint{hep-ph/0401080}.

\bibitem[{\citenamefont{Cirigliano
  et~al.}(2005{\natexlab{b}})}]{Cirigliano:2005xn}
\bibinfo{author}{\bibfnamefont{V.}~\bibnamefont{Cirigliano}}
  \bibnamefont{et~al.}, \bibinfo{journal}{JHEP} \textbf{\bibinfo{volume}{04}},
  \bibinfo{pages}{006} (\bibinfo{year}{2005}{\natexlab{b}}),
  \eprint{hep-ph/0503108}.

\bibitem[{\citenamefont{Becirevic et~al.}(2005)}]{Becirevic:2004ya}
\bibinfo{author}{\bibfnamefont{D.}~\bibnamefont{Becirevic}}
  \bibnamefont{et~al.}, \bibinfo{journal}{Nucl. Phys.}
  \textbf{\bibinfo{volume}{B705}}, \bibinfo{pages}{339} (\bibinfo{year}{2005}),
  \eprint{hep-ph/0403217}.

\bibitem[{\citenamefont{Dawson et~al.}(2006)\citenamefont{Dawson, Izubuchi,
  Kaneko, Sasaki, and Soni}}]{Dawson:2006qc}
\bibinfo{author}{\bibfnamefont{C.}~\bibnamefont{Dawson}},
  \bibinfo{author}{\bibfnamefont{T.}~\bibnamefont{Izubuchi}},
  \bibinfo{author}{\bibfnamefont{T.}~\bibnamefont{Kaneko}},
  \bibinfo{author}{\bibfnamefont{S.}~\bibnamefont{Sasaki}}, \bibnamefont{and}
  \bibinfo{author}{\bibfnamefont{A.}~\bibnamefont{Soni}},
  \bibinfo{journal}{Phys. Rev.} \textbf{\bibinfo{volume}{D74}},
  \bibinfo{pages}{114502} (\bibinfo{year}{2006}), \eprint{hep-ph/0607162}.

\bibitem[{\citenamefont{Boyle et~al.}(2008)}]{Boyle:2007qe}
\bibinfo{author}{\bibfnamefont{P.~A.} \bibnamefont{Boyle}}
  \bibnamefont{et~al.}, \bibinfo{journal}{Phys. Rev. Lett.}
  \textbf{\bibinfo{volume}{100}}, \bibinfo{pages}{141601}
  (\bibinfo{year}{2008}), \eprint{0710.5136}.

\bibitem[{\citenamefont{Lubicz et~al.}(2009)\citenamefont{Lubicz, Mescia,
  Simula, Tarantino, and Collaboration}}]{Lubicz:2009ht}
\bibinfo{author}{\bibfnamefont{V.}~\bibnamefont{Lubicz}},
  \bibinfo{author}{\bibfnamefont{F.}~\bibnamefont{Mescia}},
  \bibinfo{author}{\bibfnamefont{S.}~\bibnamefont{Simula}},
  \bibinfo{author}{\bibfnamefont{C.}~\bibnamefont{Tarantino}},
  \bibnamefont{and} \bibinfo{author}{\bibfnamefont{f.~t.~E.}
  \bibnamefont{Collaboration}} (\bibinfo{year}{2009}), \eprint{0906.4728}.

\bibitem[{\citenamefont{Cabibbo et~al.}(2003)\citenamefont{Cabibbo, Swallow,
  and Winston}}]{Cabibbo:2003cu}
\bibinfo{author}{\bibfnamefont{N.}~\bibnamefont{Cabibbo}},
  \bibinfo{author}{\bibfnamefont{E.~C.} \bibnamefont{Swallow}},
  \bibnamefont{and} \bibinfo{author}{\bibfnamefont{R.}~\bibnamefont{Winston}},
  \bibinfo{journal}{Ann. Rev. Nucl. Part. Sci.} \textbf{\bibinfo{volume}{53}},
  \bibinfo{pages}{39} (\bibinfo{year}{2003}), \eprint{hep-ph/0307298}.

\bibitem[{\citenamefont{Braaten et~al.}(1992)\citenamefont{Braaten, Narison,
  and Pich}}]{Braaten:1991qm}
\bibinfo{author}{\bibfnamefont{E.}~\bibnamefont{Braaten}},
  \bibinfo{author}{\bibfnamefont{S.}~\bibnamefont{Narison}}, \bibnamefont{and}
  \bibinfo{author}{\bibfnamefont{A.}~\bibnamefont{Pich}},
  \bibinfo{journal}{Nucl. Phys.} \textbf{\bibinfo{volume}{B373}},
  \bibinfo{pages}{581} (\bibinfo{year}{1992}).

\bibitem[{\citenamefont{Gamiz et~al.}(2003)\citenamefont{Gamiz, Jamin, Pich,
  Prades, and Schwab}}]{Gamiz:2002nu}
\bibinfo{author}{\bibfnamefont{E.}~\bibnamefont{Gamiz}},
  \bibinfo{author}{\bibfnamefont{M.}~\bibnamefont{Jamin}},
  \bibinfo{author}{\bibfnamefont{A.}~\bibnamefont{Pich}},
  \bibinfo{author}{\bibfnamefont{J.}~\bibnamefont{Prades}}, \bibnamefont{and}
  \bibinfo{author}{\bibfnamefont{F.}~\bibnamefont{Schwab}},
  \bibinfo{journal}{JHEP} \textbf{\bibinfo{volume}{01}}, \bibinfo{pages}{060}
  (\bibinfo{year}{2003}), \eprint{hep-ph/0212230}.

\bibitem[{\citenamefont{Aubin et~al.}(2004)}]{Aubin:2004fs}
\bibinfo{author}{\bibfnamefont{C.}~\bibnamefont{Aubin}} \bibnamefont{et~al.}
  (\bibinfo{collaboration}{MILC}), \bibinfo{journal}{Phys. Rev.}
  \textbf{\bibinfo{volume}{D70}}, \bibinfo{pages}{114501}
  (\bibinfo{year}{2004}), \eprint{hep-lat/0407028}.

\bibitem[{\citenamefont{Beane et~al.}(2007)\citenamefont{Beane, Bedaque,
  Orginos, and Savage}}]{Beane:2006kx}
\bibinfo{author}{\bibfnamefont{S.~R.} \bibnamefont{Beane}},
  \bibinfo{author}{\bibfnamefont{P.~F.} \bibnamefont{Bedaque}},
  \bibinfo{author}{\bibfnamefont{K.}~\bibnamefont{Orginos}}, \bibnamefont{and}
  \bibinfo{author}{\bibfnamefont{M.~J.} \bibnamefont{Savage}},
  \bibinfo{journal}{Phys. Rev.} \textbf{\bibinfo{volume}{D75}},
  \bibinfo{pages}{094501} (\bibinfo{year}{2007}), \eprint{hep-lat/0606023}.

\bibitem[{\citenamefont{Follana et~al.}(2008)\citenamefont{Follana, Davies,
  Lepage, and Shigemitsu}}]{Follana:2007uv}
\bibinfo{author}{\bibfnamefont{E.}~\bibnamefont{Follana}},
  \bibinfo{author}{\bibfnamefont{C.~T.~H.} \bibnamefont{Davies}},
  \bibinfo{author}{\bibfnamefont{G.~P.} \bibnamefont{Lepage}},
  \bibnamefont{and}
  \bibinfo{author}{\bibfnamefont{J.}~\bibnamefont{Shigemitsu}}
  (\bibinfo{collaboration}{HPQCD}), \bibinfo{journal}{Phys. Rev. Lett.}
  \textbf{\bibinfo{volume}{100}}, \bibinfo{pages}{062002}
  (\bibinfo{year}{2008}), \eprint{0706.1726}.

\bibitem[{\citenamefont{Blossier et~al.}(2009)}]{Blossier:2009bx}
\bibinfo{author}{\bibfnamefont{B.}~\bibnamefont{Blossier}} \bibnamefont{et~al.}
  (\bibinfo{year}{2009}), \eprint{0904.0954}.

\bibitem[{\citenamefont{Bazavov et~al.}(2009)}]{Bazavov:2009bb}
\bibinfo{author}{\bibfnamefont{A.}~\bibnamefont{Bazavov}} \bibnamefont{et~al.}
  (\bibinfo{year}{2009}), \eprint{0903.3598}.

\bibitem[{\citenamefont{Marciano}(2004)}]{Marciano:2004uf}
\bibinfo{author}{\bibfnamefont{W.~J.} \bibnamefont{Marciano}},
  \bibinfo{journal}{Phys. Rev. Lett.} \textbf{\bibinfo{volume}{93}},
  \bibinfo{pages}{231803} (\bibinfo{year}{2004}), \eprint{hep-ph/0402299}.

\bibitem[{\citenamefont{Wood et~al.}(1997)}]{Wood:1997zq}
\bibinfo{author}{\bibfnamefont{C.~S.} \bibnamefont{Wood}} \bibnamefont{et~al.},
  \bibinfo{journal}{Science} \textbf{\bibinfo{volume}{275}},
  \bibinfo{pages}{1759} (\bibinfo{year}{1997}).

\bibitem[{\citenamefont{Vetter et~al.}(1995)\citenamefont{Vetter, Meekhof,
  Majumder, Lamoreaux, and Fortson}}]{Vetter:1995vf}
\bibinfo{author}{\bibfnamefont{P.~A.} \bibnamefont{Vetter}},
  \bibinfo{author}{\bibfnamefont{D.~M.} \bibnamefont{Meekhof}},
  \bibinfo{author}{\bibfnamefont{P.~K.} \bibnamefont{Majumder}},
  \bibinfo{author}{\bibfnamefont{S.~K.} \bibnamefont{Lamoreaux}},
  \bibnamefont{and} \bibinfo{author}{\bibfnamefont{E.~N.}
  \bibnamefont{Fortson}}, \bibinfo{journal}{Phys. Rev. Lett.}
  \textbf{\bibinfo{volume}{74}}, \bibinfo{pages}{2658} (\bibinfo{year}{1995}).

\bibitem[{\citenamefont{Zeller et~al.}(2002)}]{Zeller:2001hh}
\bibinfo{author}{\bibfnamefont{G.~P.} \bibnamefont{Zeller}}
  \bibnamefont{et~al.} (\bibinfo{collaboration}{NuTeV}),
  \bibinfo{journal}{Phys. Rev. Lett.} \textbf{\bibinfo{volume}{88}},
  \bibinfo{pages}{091802} (\bibinfo{year}{2002}), \eprint{hep-ex/0110059}.

\bibitem[{\citenamefont{Blondel et~al.}(1990)}]{Blondel:1989ev}
\bibinfo{author}{\bibfnamefont{A.}~\bibnamefont{Blondel}} \bibnamefont{et~al.},
  \bibinfo{journal}{Z. Phys.} \textbf{\bibinfo{volume}{C45}},
  \bibinfo{pages}{361} (\bibinfo{year}{1990}).

\bibitem[{\citenamefont{Allaby et~al.}(1986)}]{Allaby:1986pd}
\bibinfo{author}{\bibfnamefont{J.~V.} \bibnamefont{Allaby}}
  \bibnamefont{et~al.} (\bibinfo{collaboration}{CHARM}),
  \bibinfo{journal}{Phys. Lett.} \textbf{\bibinfo{volume}{B177}},
  \bibinfo{pages}{446} (\bibinfo{year}{1986}).

\bibitem[{\citenamefont{McFarland et~al.}(1998)}]{McFarland:1997wx}
\bibinfo{author}{\bibfnamefont{K.~S.} \bibnamefont{McFarland}}
  \bibnamefont{et~al.} (\bibinfo{collaboration}{CCFR}), \bibinfo{journal}{Eur.
  Phys. J.} \textbf{\bibinfo{volume}{C1}}, \bibinfo{pages}{509}
  (\bibinfo{year}{1998}), \eprint{hep-ex/9701010}.

\bibitem[{\citenamefont{Vilain et~al.}(1994)}]{Vilain:1994qy}
\bibinfo{author}{\bibfnamefont{P.}~\bibnamefont{Vilain}} \bibnamefont{et~al.}
  (\bibinfo{collaboration}{CHARM-II}), \bibinfo{journal}{Phys. Lett.}
  \textbf{\bibinfo{volume}{B335}}, \bibinfo{pages}{246} (\bibinfo{year}{1994}).

\bibitem[{\citenamefont{Coll.}(2003)}]{:2003ih}
\bibinfo{author}{\bibnamefont{Coll.}} (\bibinfo{collaboration}{LEP})
  (\bibinfo{year}{2003}), \eprint{hep-ex/0312023}.

\bibitem[{\citenamefont{Coll.}(2006)}]{:2005ema}
\bibinfo{author}{\bibnamefont{Coll.}} (\bibinfo{collaboration}{ALEPH}),
  \bibinfo{journal}{Phys. Rept.} \textbf{\bibinfo{volume}{427}},
  \bibinfo{pages}{257} (\bibinfo{year}{2006}), \eprint{hep-ex/0509008}.

\bibitem[{\citenamefont{Abbiendi et~al.}(2004)}]{Abbiendi:2003dh}
\bibinfo{author}{\bibfnamefont{G.}~\bibnamefont{Abbiendi}} \bibnamefont{et~al.}
  (\bibinfo{collaboration}{OPAL}), \bibinfo{journal}{Eur. Phys. J.}
  \textbf{\bibinfo{volume}{C33}}, \bibinfo{pages}{173} (\bibinfo{year}{2004}),
  \eprint{hep-ex/0309053}.

\bibitem[{\citenamefont{Achard et~al.}(2004)}]{Achard:2004zw}
\bibinfo{author}{\bibfnamefont{P.}~\bibnamefont{Achard}} \bibnamefont{et~al.}
  (\bibinfo{collaboration}{L3}), \bibinfo{journal}{Phys. Lett.}
  \textbf{\bibinfo{volume}{B600}}, \bibinfo{pages}{22} (\bibinfo{year}{2004}),
  \eprint{hep-ex/0409016}.

\bibitem[{\citenamefont{Abazov et~al.}(2004)}]{Abazov:2003sv}
\bibinfo{author}{\bibfnamefont{V.~M.} \bibnamefont{Abazov}}
  \bibnamefont{et~al.} (\bibinfo{collaboration}{CDF}), \bibinfo{journal}{Phys.
  Rev.} \textbf{\bibinfo{volume}{D70}}, \bibinfo{pages}{092008}
  (\bibinfo{year}{2004}), \eprint{hep-ex/0311039}.

\bibitem[{\citenamefont{Antonelli et~al.}(2009)}]{Antonelli:2009ws}
\bibinfo{author}{\bibfnamefont{M.}~\bibnamefont{Antonelli}}
  \bibnamefont{et~al.} (\bibinfo{year}{2009}), \eprint{0907.5386}.

\bibitem[{\citenamefont{Barbieri and Strumia}(1999)}]{Barbieri:1999tm}
\bibinfo{author}{\bibfnamefont{R.}~\bibnamefont{Barbieri}} \bibnamefont{and}
  \bibinfo{author}{\bibfnamefont{A.}~\bibnamefont{Strumia}},
  \bibinfo{journal}{Phys. Lett.} \textbf{\bibinfo{volume}{B462}},
  \bibinfo{pages}{144} (\bibinfo{year}{1999}), \eprint{hep-ph/9905281}.

\end{thebibliography}

\end{document}